\documentclass[12pt,letterpaper]{article}
\pdfoutput=1

\usepackage{color}
\usepackage{cite}
\usepackage{slashed}
\definecolor{refkey}{rgb}{1,0,0}
\definecolor{labelkey}{rgb}{0,0,1}
\usepackage{amsfonts,amssymb}
\usepackage{amsfonts,amssymb,amsmath}
\usepackage{MnSymbol,wasysym}
\usepackage{esint}
\usepackage{rotating}

\usepackage{graphicx}
\usepackage{caption}
\usepackage{subcaption}
\usepackage{booktabs}

\usepackage{listings}

\usepackage{hyperref}

\numberwithin{equation}{section}

\newcommand{\intxi}{\ensuremath{\int_0^\pi\!\!\!\!\!\! d\xi\,}}

\newcommand{\be}{\begin{equation}}
\newcommand{\ee}{\end{equation}}
\newcommand{\ben}{\begin{displaymath}}
\newcommand{\een}{\end{displaymath}}
\newcommand{\bea}{\begin{equation}\begin{aligned}}
\newcommand{\eea}{\end{aligned}\end{equation}}
\newcommand{\bean}{\begin{eqnarray*}}
\newcommand{\eean}{\end{eqnarray*}}

\def\a {\alpha}

\newcommand{\eg}{{\it e.g.}}
\newcommand{\ie}{{\it i.e.}}

\newcommand{\commentout}[1]{}

\usepackage{amsmath}

\newcommand{\beq}{\begin{equation}}
\newcommand{\eeq}{\end{equation}}
\newcommand{\beqr}{\begin{displaymath}}
\newcommand{\eeqr}{\end{displaymath}}
\newcommand{\beqa}{\begin{eqnarray}}
\newcommand{\eeqa}{\end{eqnarray}}
\newcommand{\beqar}{\begin{eqnarray*}}
\newcommand{\eeqar}{\end{eqnarray*}}

\newcommand{\non}{\nonumber}

\newcommand{\cF}{{\cal F}}
\newcommand{\cK}{{\cal K}}

\newcommand{\cC}{{\cal C}}

\newcommand{\half}{\ensuremath{\frac{1}{2}}}

\renewcommand{\Re}{\ensuremath{\mathrm{Re}}}
\renewcommand{\Im}{\ensuremath{\mathrm{Im}}}

\newcommand{\mmax}[1]{\ensuremath{\underset{#1}{\max}}}
\newcommand{\mmin}[1]{\ensuremath{\underset{#1}{\min}}}

\newcommand{\Mreg}{\ensuremath{M_{\mbox{reg}}}}

\newcommand{\cone}{\ensuremath{\mathrm{C}}}

\newcommand{\intxy}{\ensuremath{\int_4^\infty\!\!\!\!\!\! dx \int_4^\infty\!\!\!\!\!\! dy\,}}
\newcommand{\intx}{\ensuremath{\int_4^\infty\!\!\!\!\!\! dx\,}}

\newcommand{\intP}{\ensuremath{\int_{-1}^{+1}\!\!\!\!\!\! d\mu\, P_\ell(\mu)\,}}
\newcommand{\intls}{\ensuremath{\sum_\ell\int_4^{\infty}\!\!\!\!\!\! ds\,}}
\newcommand{\intlsI}{\ensuremath{\sum_{I,\ell}\int_4^{\infty}\!\!\!\!\!\! ds\,}}
\newcommand{\intxyi}{\ensuremath{\int_{-\infty}^{+\infty}\!\!\!\!\!\! dx \int_{-\infty}^{+\infty}\!\!\!\!\!\! dy\,}}
\newcommand{\intst}{\ensuremath{\int_4^\infty\!\!\!\!\!\! ds \int_{4-s}^0\!\!\!\!\!\! dt\,}}

\newcommand{\Qx}{\ensuremath{Q_\ell\left(1+\frac{2x}{s-4}\right)}}
\newcommand{\Qy}{\ensuremath{Q_\ell\left(1+\frac{2y}{s-4}\right)}}
\newcommand{\sumeven}{\sum_{\ell\,\mathrm{even}}}

\newcommand{\intlevens}{\ensuremath{\sum_{\ell\,\mathrm{even}}\int_4^{\infty}\!\!\!\!\!\! ds\,}}

\begin{document}

\title{\Large \bf S-matrix bootstrap in 3+1 dimensions:\\ regularization and dual convex problem}

\author{
	Yifei He$^\text{1,2}$,
	Martin Kruczenski$^\text{3}$ \thanks{E-mail: \texttt{yifei.he@phys.ens.fr, markru@purdue.edu.}} \\
[2.0mm]
$^1$ \small Institut Philippe Meyer, \'Ecole Normale Sup\'erieure,\\
\small Universit\'e PSL, 24 rue Lhomond, F-75231 Paris, France \\
$^2$ \small Universit\'e Paris-Saclay, CNRS, CEA, \\
\small Institut de Physique Th\'eorique,
91191, Gif-sur-Yvette, France \\
$^3$ \small Department of Physics and Astronomy and PQSEI\thanks{Purdue Quantum Science and Engineering Institute}, \\
\small Purdue University, West Lafayette, IN 47907, USA.}

\date{}

\maketitle

\begin{abstract}
The S-matrix bootstrap maps out the space of S-matrices allowed by analyticity, crossing, unitarity, and other constraints. For the $2\rightarrow 2$ scattering matrix $S_{2\rightarrow 2}$ such space is an infinite dimensional convex space whose boundary can be determined by maximizing linear functionals.  On the boundary interesting theories can be found, many times at vertices of the space. Here we consider $3+1$ dimensional theories and focus on the equivalent dual convex minimization problem that provides strict upper bounds for the regularized primal problem and has interesting practical and physical advantages over the primal problem. Its variables are dual partial waves $k_\ell(s)$ that are free variables, namely they do not have to obey any crossing, unitarity or other constraints. Nevertheless they are directly related to the partial waves $f_\ell(s)$, for which all crossing, unitarity and symmetry properties result from the minimization. Numerically, it requires only a few dual partial waves, much as one wants to possibly match experimental results. We consider the case of scalar fields which is related to pion physics. 
\end{abstract}

\clearpage

\tableofcontents

\newpage




\section{Introduction and Summary}
\label{intro}

 Recently, new insights were found on the old idea \cite{Eden:1966dnq,chew1966analytic} of solving the S-matrix directly from its analytic structure, symmetries, crossing and unitarity. Although these constraints do not uniquely determine the S-matrix, combining them with a convex maximization program leads to bounds on couplings between particles and their bound states. Furthermore, it was observed that well-known theories such as a subsector of the sine-Gordon model saturate such bounds \cite{Paulos:2016fap,Paulos:2016but}. Similar programs can be carried out for 3+1 dimensional theories \cite{Paulos:2017fhb}, and multiple amplitudes \cite{Homrich:2019cbt}. 
 Even in the case where the theory does not have bound states similar ideas can be applied. An interesting example is the 1+1 d $O(N)$ non-linear sigma model \cite{Polyakov:1975rr} which describes an asymptotically free theory with a dynamically developed mass gap in the infrared. It was exactly solved in \cite{Zamolodchikov:1977nu,Zamolodchikov:1978xm} and more recently revisited with the S-matrix bootstrap approach in \cite{He:2018uxa,Cordova:2018uop,Paulos:2018fym}. In particular, in \cite{He:2018uxa} it was argued that one should focus on mapping out the space of allowed S-matrices under the given constraints. From that perspective, the $O(N)$ non-linear sigma model lies at a special point -- a vertex -- in the space of allowed theories, and maximizing a linear functional in a convex space generically leads to a vertex. This idea was made more manifest later in \cite{Cordova:2019lot} where a section of the space was plotted with a clear vertex at the NLSM. Further work on other models \cite{Bercini:2019vme} showed that sometimes full regions of the boundary correspond to interesting theories if such theories have free parameters. More recently, the study of reflection matrices \cite{Kruczenski:2020ujw} showed that vertices appear at points where resonances, namely poles in the second sheet cross into the physical region becoming bound states. Since bound states are not allowed, the S-matrix changes form and a discontinuity in the derivative appears.
 
For the $3+1$ dimensional case considered here, besides the previously mentioned references, the most relevant ones are \cite{Guerrieri:2018uew, Guerrieri:2020bto} where the S-matrix bootstrap is applied to pion dynamics as originally envisioned \cite{Eden:1966dnq,chew1966analytic}.
Various other ideas have been discussed in the context of the S-matrix bootstrap and similar methods applied to gapped theories\cite{Doroud:2018szp,EliasMiro:2019kyf,Karateev:2019ymz,Correia:2020xtr,Bose:2020shm,Komatsu:2020sag,Bose:2020cod,Hebbar:2020ukp,Karateev:2020axc,Tourkine:2021fqh,Guerrieri:2021ivu,Anderson:2016rcw,Elvang:2020lue,Huang:2020nqy}.
 
 \medskip
 
 In this work we concentrate on the dual problem proposed in \cite{Cordova:2019lot} and further studied in \cite{Guerrieri:2020kcs,pollica}. Thinking in terms of the allowed space of S-matrices, the primal approach provides an interior region that becomes larger as the numerics is improved, eventually approaching the full allowed space. On the other hand, the dual problem provides an exterior region that shrinks onto the allowed space of S-matrices as the numerics improves. This is particularly useful in $3+1$ dimensions where the problem is numerically difficult, since it allows to bracket the boundary between the interior and the exterior. Moreover, we observed that the dual problem requires only a few partial waves much in the same way as the results of experiments where only a few partial waves are available. In summary, the dual problem as described here provides a complementary and in many ways very useful approach to the problem of mapping out the allowed space of S-matrices. 
 
 The construction of the dual problem requires several steps. First we need to define the primal problem. For that we use the Mandelstam representation which assumes maximal analyticity and enforces crossing properties for the amplitudes. The unitarity constraints are then imposed on the partial waves with definite angular momentum and isospin. It turns out, however, that the price to pay for a parameterization that has manifest analyticity and crossing is that, it contains fluctuations of the variables that barely modify the partial waves. Therefore, roughly speaking, they are not constrained by unitarity and they do not contribute to the physical partial waves. We solved this problem by regularizing the primal problem, namely putting a bound -- a regulator -- on the variation of the primal variables. A large enough regulator allows for the amplitudes we need, while suppressing unwanted large fluctuations of irrelevant variables. In addition, the regularization also leads to a well-defined dual problem. Alternatively, one could parameterize the partial waves as analytic functions, but in that case crossing is difficult to impose since it mixes all angular momenta. We have not explored this last possibility further. 
 
\begin{figure}
 	\centering
 	\includegraphics[width=\textwidth]{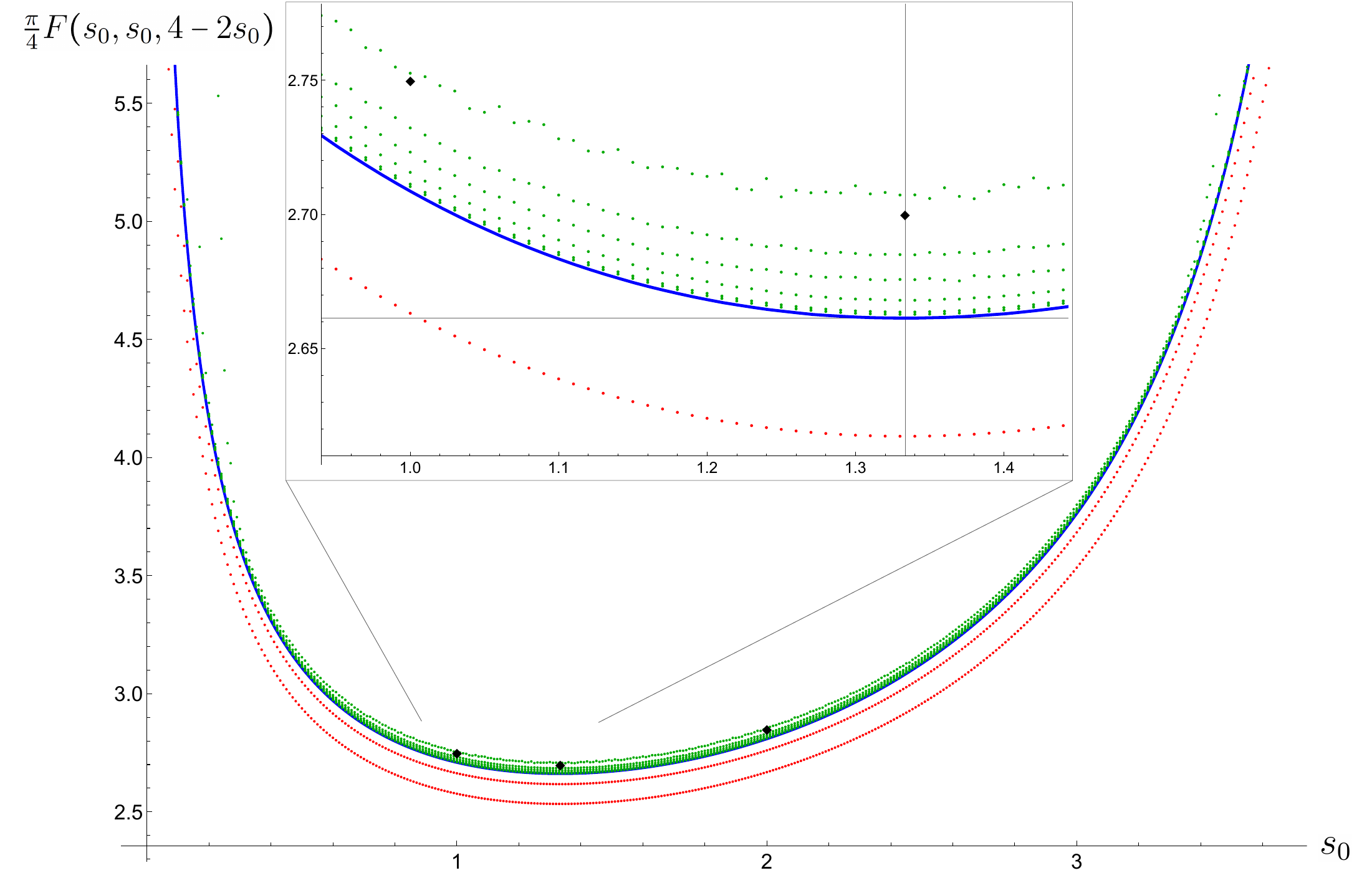}
 	\caption{The maximum of the amplitudes along $s_0=t_0$ in the (large) Mandelstam triangle. The green points are the dual (upper bounds), the red are the primal (lower bounds for the maximum). Both were run without assuming the existence of a pole at threshold. The blue line represents the primal problem once a pole at threshold is included. The black diamonds represent the pioneering results of Lopez and Mennessier \cite{LM}. The inset shows more clearly the results near the symmetric point which is the minimum of the blue curve where $\frac{\pi}{4}F(\frac{4}{3},\frac{4}{3},\frac{4}{3})\simeq2.6613$. The factor $\pi/4$ is to match the normalization in the literature, \eg\ \cite{Paulos:2017fhb}.}
 	\label{fig3}
\end{figure}

 Once the primal problem is regularized, it is straightforward to define the dual problem using the tools of conic optimization. This leads to upper bounds on the space of S-matrices allowed by the constraints. As an example, we can see in figure \ref{fig3} how the maximum (blue curve) is bracketed between the red curves (primal) and the green curves (dual). In this case, the blue curve was obtained by assuming the existence of a pole at threshold whereas the red and green curves did not make such assumption since, in general, we do not know the pole structure of the optimal amplitude. Furthermore, as in 2d, it turns out that one can derive the dual problem just using analytic properties of the amplitudes. For that purpose we introduce what we call the generalized double dispersion relations \eqref{formula} that allows us to relate the values of the amplitude at an unphysical point to the values in the physical region. We do this by defining a dual amplitude (see for example \eqref{Kdisp}) through a double dispersion relation with support in the physical region. After minimization, the dual amplitude contains all the information necessary to extract the partial waves in the energy range where they saturate unitarity. The construction allows for the possibility of unitarity unsaturation but we did not find that property in the numerical solutions we obtained. It will be interesting to explore this idea further. 
  
  \medskip
  
  This paper is organized as follows. In the next section, we review how conic optimization can be used to map out the space of allowed S-matrices. In the following section, we describe first how  regularizing the primal problem eliminates irrelevant fluctuations of the primal variables while at the same time provides a well-defined dual problem. In section \ref{sec4} we rederive the dual problem using arguments from complex analysis, where we introduce the generalized dispersion relation \eqref{formula} and the dual amplitudes. These are analytic functions of two variables that contain the same physical information as the amplitude to the extent that the partial waves can be extracted from it where unitarity constraints are saturated. Its double spectral density has support in the physical region instead of the Mandelstam region. Although the subsequent numerical tests focus on the single flavor case, we have also included the formulation with $O(N)$ global symmetry in sections \ref{ONp}, \ref{ONd} and \ref{ONcomplex}, which may be skipped for readers who want to focus on the general logic.
  In section \ref{sec5} we describe two different numerical implementations of the primal and dual problems, one based on interpolation points and the other using a basis of functions. We illustrate these methods by doing some simple numerical computations in section \ref{sec6} and leave a full numerical exploration of the method for upcoming work. We give our conclusions in section \ref{con} and collect a few useful formulas in an appendix.
  
\section{Physical problem as a primal problem in conic optimization}
\label{sec2}

In this paper we consider $2\rightarrow 2$ scattering of scalar particles because of its simplicity. Although we do not assume any particular identity for the particles or form of the interaction, it is useful to describe this process in the language of pion scattering because it is a very well-studied example. This type of system has already been studied with the S-matrix bootstrap in \cite{Paulos:2017fhb,Guerrieri:2018uew,Guerrieri:2020bto,Correia:2020xtr,Bose:2020shm}. In this paper we concentrate on the formulation of the dual problem. 

\subsection{Single pion scattering}
\begin{figure}
	\centering
	\includegraphics[width=0.8\textwidth]{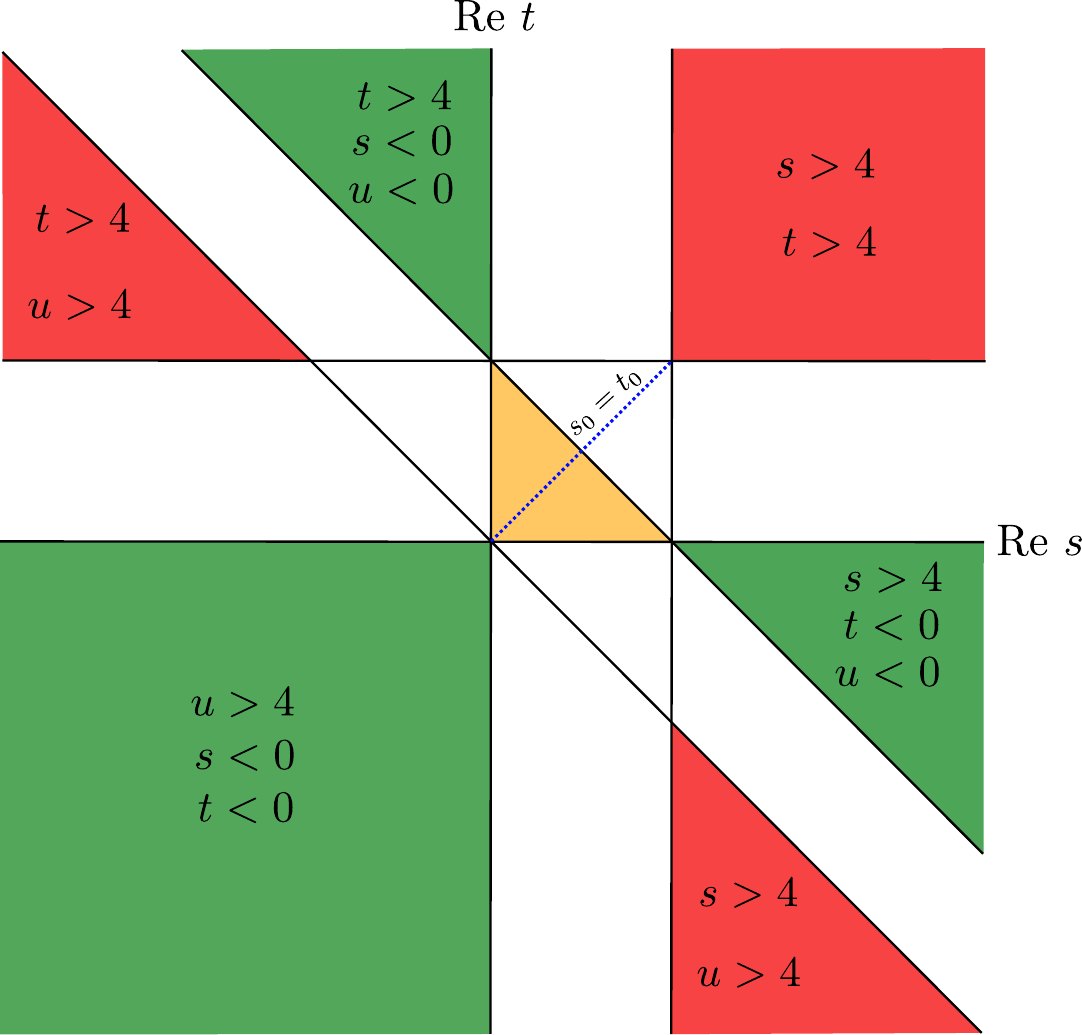}
	\caption{Mandelstam diagram in the real ($s$, $t$, $u$) plane. The red regions are the Mandelstam regions where the amplitude spectral densities $\rho^{st}$ and $\rho^{su}$ in \eqref{a106} have support. The green regions are the physical regions in the $s$, $t$ and $u$ channels where the dual amplitudes \eqref{Kdualsp} and \eqref{KdualON} have double jumps. The triangles in the middle are the Mandelstam triangles where we usually evaluate the amplitude for maximization and the diagonal blue line $s_0=t_0$ is where we maximize the amplitude to produce fig.\ref{fig3}. }
	\label{regions}
\end{figure}

The simplest example is single pion scattering, namely a scalar particle of mass $m^2$ that we set to $m^2=1$. The amplitude for the process
\beq
\pi(p_1) + \pi(p_2) \rightarrow \pi(p_3) +\pi(p_4), 
\label{a1}
\eeq
is a function of the Mandelstam variables $s=(p_1+p_2)^2$, $t=(p_1-p_3)^2$, $u=(p_1-p_4)^2$ with $s+t+u=4$. Its analytic and crossing properties are captured by the Mandelstam representation that can be written as\footnote{Note that in using the Mandelstam representation we have assumed maximal analyticity which is not required to formulate the primal and dual problems as discussed in section \ref{singlepi}. However, such assumption is important for the numerical implementation of the bootstrap problem used in this paper and to compare with the primal problem in \cite{Paulos:2017fhb}.}
\beq
 F(s,t,u) = f_0 + \intx \cK(s,t,u;x)  \sigma(x) + \intxy \cK(s,t,u;x,y)\, \rho(x,y) 
 \label{a2}
\eeq 
with the kernels
\begin{subequations}
	\beqa
	\cK(s,t,u;x) &=& \frac{1}{\pi} \left[\frac{1}{x-s} + \frac{1}{x-t}+\frac{1}{x-u}\right]  \label{a3}\\
	\cK(s,t,u;x,y) &=&  \frac{1}{2\pi^2}\left[\frac{1}{(x-s)(y-t)}+\frac{1}{(x-s)(y-u)} + \frac{1}{(x-u)(y-t)}\right] \nonumber\\
	&&+(x\leftrightarrow y)  \label{a3b}
	\eeqa
\end{subequations}
The amplitude $F(s,t,u)$ has double jumps on the Mandelstam regions depicted in red in fig.\ref{regions}. The amplitude is crossing symmetric under $s\leftrightarrow t \leftrightarrow u$, and only the symmetric part of $\rho$ contributes so we can take $\rho(x,y)=\rho(y,x)$. The physical partial waves $f_\ell(s)$ are defined as
\beq
f_\ell(s) = \frac{1}{4} \intP F\left(s^+,t,u\right) \label{hldef}\\\
\eeq
which are non-zero only for even $\ell$. In the integral \eqref{hldef}, $t$ and $u$ are to be considered as functions of $\mu$ and $s$ according to
\begin{subequations}
	\beqa
	t(\mu,s) &=& -\frac{(s-4)(1-\mu)}{2} \label{a4}\\
	u(\mu,s) &=& 4-s-t(\mu,s)=t(-\mu,s).  \label{a5}
	\eeqa
\end{subequations}
From now on, when integrating over Legendre polynomials of $\mu$ we always assume $t=t(\mu,s)$, $u=u(\mu,s)$ according to (\ref{a4}), (\ref{a5}). This implies that, to compute the partial waves, we only need to know the amplitude in the physical region depicted in green in fig.\ref{regions} (or any of its crossing conjugates since the amplitude is crossing symmetric). It is convenient to define rescaled partial waves $h_\ell(s)$ and the S-matrices $S_\ell(s)$:
\beqa
 h_\ell(s) &=& \pi\sqrt{\frac{s-4}{s}} f_\ell(s)  \label{a6} \\
S_\ell(s) &=& 1+i\,h_\ell(s) =e^{2i\delta_\ell} \label{Sldef}
\eeqa
where $\delta_\ell$ are the (possibly complex) phase-shifts. The unitarity condition is 
\beq
 |S_\ell(s)|\le 1, \forall \ell\in\mathbb{Z}_{\ge 0}, \ \ \ s\in\mathbb{R}_{\ge 4} \ \ \ \Leftrightarrow \ \ |h_\ell(s)|^2 \le 2\, \Im h_\ell(s) 
 \label{unitarity1}
\eeq
Notice that in the physical region $s>4$, $4-s<t<0$, eq.(\ref{hldef}) can be inverted as
\beq
 F(s^+,t,u) = \frac{2}{\pi}\sqrt{\frac{s}{s-4}} \sumeven (2\ell+1) h_\ell(s) P_\ell\left(1+\frac{2t}{s-4}\right)
 \label{Hinv}
\eeq 
where the sum is over even $\ell$ and $s^+$ indicates we evaluate the amplitude on the real axis above the cut $[4,\infty)$.  

 Given the general constraints of crossing, unitarity and analyticity as given by the Mandelstam representation, we want to map out the space of allowed amplitudes. One way to do that is to start from a point, \eg\ the free theory $S_\ell=1$, pick a direction in the space of S-matrices and move along that direction until one of the constraints is violated. This is equivalent to maximizing a linear functional. The space of allowed S-matrices is of course infinite dimensional, but we can concentrate in a lower dimensional subspace. Since unitarity is a convex constraint and we impose linear conditions, the problem is a conic maximization problem \cite{conic1} that can be solved with standard methods \cite{cvx,gb08,boyd2004convex}. A standard functional that we can choose is 
 \beq
  \cF_P = F(s_0,t_0,u_0) \label{a7}
 \eeq
where $(s_0,t_0,u_0)$ is a point in the Mandelstam triangle (see fig. \ref{regions}) where the amplitude is real. So the problem is to find $(f_0,\sigma(x),\rho(x,y)=\rho(y,x))$ that maximizes $\cF_P$ under the unitarity constraint. In convex maximization, an important role is played by the dual problem that we define later and is the main subject of our paper.

\subsection{$O(N)$ case}\label{ONp}
Now we consider a scalar theory with $N$ species of particles and $O(N)$ symmetry. For the actual pions $N=3$. The $2\rightarrow 2$ amplitude 
\beq
\pi_a(p_1) + \pi_b(p_2) \rightarrow \pi_c(p_3) +\pi_d(p_4), 
\label{b2}
\eeq
is now
\beq
 F_{ab,cd} = A(s,t,u) \delta_{ab} \delta_{cd} + A(t,s,u) \delta_{ac} \delta_{bd} + A(u,t,s) \delta_{ad}\delta_{bc} 
 \label{a8}
\eeq
with $A(s,t,u)=A(s,u,t)$ and $a,b,c,d\ldots=1\ldots N$. The expression satisfies crossing and isospin symmetry but we still need to impose the unitarity constraint. For a well-defined isospin ($I=0,1,2$) in the $s$-channel we get:
\beq
 F_{ab,cd} = \frac{1}{N} F^{I=0} \delta_{ab}\delta_{cd} + \frac{1}{2} F^{I=1} (\delta_{ac}\delta_{bd}-\delta_{ad}\delta_{bc}) + \half F^{I=2} (\delta_{ac}\delta_{bd}+\delta_{ad}\delta_{bc}-\frac{2}{N}\delta_{ab}\delta_{cd})  
\label{a9}
\eeq 
with 
\begin{subequations}\label{a10}
	\beqa	
	F^{I=0}(s,t,u) &=& NA(s,t,u)+A(t,s,u)+A(u,t,s) \\ 
	F^{I=1}(s,t,u) &=& A(t,s,u)-A(u,t,s) \\ 
	F^{I=2}(s,t,u) &=& A(t,s,u)+A(u,t,s) 
	\eeqa
\end{subequations}
The Mandelstam representation for the scattering amplitude, including subtractions, is now ($\alpha=1,2$)
\beq
 A(s,t,u) = f_0 + \intx  \cK_\a(s,t,u;x)  \sigma_\a(x) + \intxy \cK_\a(s,t,u;x,y)\, \rho_\a(x,y) 
\label{a11}
\eeq
with kernels
\begin{subequations}
	\beqa
	\cK_1(s,t,u;x) &=& \frac{1}{\pi} \left[ \frac{1}{x-s}\right]   \label{a12} \\
	\cK_2(s,t,u;x) &=& \frac{1}{\pi} \left[ \frac{1}{x-t}+\frac{1}{x-u}\right] \label{a13}\\
	\cK_1(s,t,u;x,y) &=&  \frac{1}{\pi^2}\left[\frac{1}{(x-s)(y-t)}+\frac{1}{(x-s)(y-u)}\right] \label{a14} \\
	\cK_2(s,t,u;x,y) &=&  \frac{1}{2\pi^2}\left[\frac{1}{(x-t)(y-u)}+ \frac{1}{(x-u)(y-t)}\right] \label{a15}
	\eeqa
\end{subequations}
The amplitude $A(s,t,u)$ has double jumps in the Mandelstam regions depicted in red in fig. \ref{regions}.
The variables now are $\sigma_{\a=1,2}$ and $\rho_{\a=1,2}(x,y)$ where $\rho_2(x,y)=\rho_2(y,x)$ and the rescaled partial waves $h^I_\ell(s)$ are given by
\beq
 h_\ell^I(s) = \frac{\pi}{4} \sqrt{\frac{s-4}{s}} \intP F^I(s^+,t,u)\, d\mu  \label{a16}
\eeq
where the even $\ell$'s are non-vanishing for $I=0,2$ and the odd ones are non-vanishing for $I=1$. Keep in mind that $t$ and $u$ are functions of $\mu$, $s$ according  to (\ref{a4}), (\ref{a5}). We can also compute a Mandelstam representation for the isospin channels:
\beq
  F^I(s,t,u) = f^I + \intx  \cK^I_\a(s,t,u;x)  \sigma_\a(x) + \intxy \cK^I_\a(s,t,u;x,y)\, \rho_\a(x,y) 
\eeq
where 
\beq
 f^0 = (N+2) f_0, \ \ f^1=0, \ \ \ f^2 = 2f_0 \ .
\eeq
The kernels follow the pattern in \eqref{a10}:
\begin{subequations}
	\beqa
	\cK^{I=0}_\alpha(s,t,u;x) &=& N\cK_\alpha(s,t,u;x)+\cK_\alpha(t,s,u;x)+\cK_\alpha(u,t,s;x) \label{c1}\\ 
	\cK^{I=1}_\alpha(s,t,u;x) &=& \cK_\alpha(t,s,u;x)-\cK_\alpha(u,t,s;x)  \label{c2}\\ 
	\cK^{I=2}_\alpha(s,t,u;x) &=& \cK_\alpha(t,s,u;x)+\cK_\alpha(u,t,s;x)  \label{c3}
	\eeqa
\end{subequations}
and the same for the double dispersion kernels.

The unitarity constraint is:
\beq
 |S_\ell^I(s)| = |1+ih^I_\ell(s)| \le 1, \ \ \ s\in\mathbb{R}_{\ge 4}  \label{a17}
\eeq 
Under these conditions we want to maximize a linear functional $\cF_P$ of the amplitude. A standard example is
\beq
\cF_P = \sum_I n_I F^I(s_0,t_0,u_0)  \label{a18}
\eeq
for a point $(s_0, t_0, u_0)$ inside the Mandelstam triangle where the amplitude is real and $n_I$ are constants. We can also take a general linear functional of the partial waves
\beq
 \cF_P = \intlsI \Re\left[c_{I,\ell}(s) h^I_\ell(s)\right]   \label{a19}
\eeq
for some coefficients $c_{I,\ell}$ of which only a finite number we choose to be non-zero. Such functional has a feasible dual problem without the need for regularization, and it can be useful if we believe that the theory we are interested in maximizes certain partial waves. 

\subsection{Primal problem}

 Both problems can be summarized in the language of convex optimization by using a compact notation as follows. Define a set of variables $\alpha_n$ where $n$ denotes a discrete or continuous index. From \eqref{a2} and \eqref{a11}, we can read
 \begin{equation}\label{alphan}
 \alpha_n=\begin{cases}
\big(f_0,\sigma(x),\rho(x,y)\big),&\text{single pion},\\
\big (f_0, \sigma_{\a=1,2}(x), \rho_{\a=1,2}(x,y)\big),&O(N)
 \end{cases}
 \end{equation}
 A sum over $n$ is understood as integration when the label is continuous. For example, for the single pion and using \eqref{a2}, the primal functional \eqref{a7} can be written as
\beq
\begin{aligned}
	 \cF_P &= F(s_0,t_0,u_0)\\
	  &= \alpha_n a_n \\
	  &= a_0 f_0 + \intx a(x)  \sigma(x) + \intxy a(x,y)\, \rho(x,y)     \label{a20}
\end{aligned}
\eeq 
Comparing with \eqref{a2}, we can extract the coefficients $a_n$ for the maximization of $\cF_P = F(s_0,t_0,u_0)$:
\begin{subequations} \label{a21f}
	\beqa
	a_0 &=& 1 \label{a21}\\
	a(x) &=& \cK(s_0,t_0,u_0;x) \label{a22}\\
	a(x,y) &=& \cK(s_0,t_0,u_0;x,y) \label{a23}
	\eeqa 
\end{subequations}
Instead, if we wanted to minimize $\cF_P = F(s_0,t_0,u_0)$, this is equivalent to maximizing $\mathcal{F}_P=-F(s_0,t_0,u_0)$ and therefore we would take the $a_n$ in \eqref{a21f} with the opposite sign. Other choices of $a_n$ would correspond to other functionals, for example those in \eqref{a19}. 
 Thus, for any choice of functional, the primal problem can be stated as
\begin{equation}
\begin{aligned}
 \mmax{\alpha_n}\;\;\Big\{\cF_P&=a_n \alpha_n\Big\}, \ \\
 s.t.  \;\; h^I_\ell(s) &= \alpha_n h^I_{\ell n}(s), \;\; \\
 |S^I_\ell(s)| &= |1+ih^I_\ell(s)| \le 1
\label{a24}
\end{aligned}
\end{equation} 
where for the single pion $I=0$, $\ell $ even, and for the $O(N)$ case $I=0,2$, $\ell$ even and $I=1$, $\ell$ odd. In \eqref{a24}, we have used the compact notation $h^{I}_{\ell n}(s)$ which gives the partial waves as linear combinations of the variables $\alpha_n$. For example, in the case of single pion, we have the following $h^I_{\ell n}(s)$ for the variables \eqref{alphan}
\begin{subequations}\label{hls}
	\beqa
	h_{\ell,0} &=& \frac{\pi}{2} \sqrt{\frac{s-4}{s}} \delta_{\ell 0} \label{a25}\\
	h_{\ell}(x;s) &=& \frac{\pi}{4} \sqrt{\frac{s-4}{s}} \intP \cK(s^+,t,u;x) \label{a26}\\
	&=& \frac{2}{\sqrt{s(s-4)}} \left[\frac{1}{4}\frac{s-4}{x-s^+}\delta_{\ell 0} + \Qx \right] \label{a27} \nonumber \\
	h_{\ell}(x,y;s) &=& \frac{\pi}{4} \sqrt{\frac{s-4}{s}} \intP \cK(s^+,t,u;x,y) \label{a28}\\
	&=& \frac{1}{\pi}\frac{1}{\sqrt{s(s-4)}} \left[\frac{1}{x-s^+}+\frac{1}{s-4+x+y}\right] \Qy  +(x\leftrightarrow y)  \label{a29} \non
	\eeqa
\end{subequations}
where we used the integrals in the appendix to write the partial waves in terms of the Legendre functions $Q_\ell$ as in the Froissart-Gribov form (see \eg\ \cite{Eden:1966dnq}) Similar expressions can be obtained for the $O(N)$ case. The values of $h^I_{\ell n}(s)$ in \eqref{hls} follow from choosing the Mandelstam representation \eqref{a2} for the amplitude that contains the assumption of maximal analyticity as used later in the paper for doing the numerical calculations. The formulation of the primal and the dual done in the next section is valid for any other values of $h^I_{\ell n}(s)$ possible reflecting other analyticity properties.

\section{Regularized primal problem and its dual}
\label{sec3}

 In this section we discuss the need and also practicality of regularizing the primal problem by putting an upper bound on the norm of the double spectral density $\rho(x,y)$ in \eqref{a2} or \eqref{a11}. This need arises from two different but related issues. One is that highly oscillating functions added to $\rho(x,y)$ make little difference in the physical amplitude, the other is that the dual problem is weakly infeasible\footnote{Namely, if we think the dual constraints as a vector, that the problem is infeasible means that it cannot be set to zero as required. Weakly infeasible means that its norm can be made as small as one wants \cite{conic1}}.  These problems are not related to the Mandelstam representation but to the fact that the unitarity condition is evaluated in the physical region $s>4$, $4-s<t<0$ which is not in the support of the double spectral density, namely the boundary of the analyticity region. For example, we can alternatively do as in \cite{Paulos:2017fhb} and map the analyticity region to unit disks for each Mandelstam variable using
\beq
 z_s = \frac{2-\sqrt{4-s}}{2+\sqrt{4-s}}, 
\label{b3}
\eeq 
with $|z_s|\le 1$ and the same for $t$ and $u$.  Then, in the physical region $|z_s|=1$ but $|z_t|<1$, $|z_u|<1$. If the single pion amplitude is written as a polynomial of some given degree $M$ as
\beq
 F(s,t,u) = \sum_{n,m=0}^{M} \rho_{nm}(z_s^n z_t^m+z_s^n z_u^m + z_t^n z_u^m) 
\label{b4}
\eeq 
 then the coefficients $\rho_{nm}$ with both $n$ and $m$ large affect the value of the polynomial very little in the physical region where $|z_t|<1$ and $|z_u|<1$. Therefore the unitarity condition does not constrain those coefficients because they do not affect the physical amplitudes and therefore they can become large. However since they do not affect the physical amplitudes they are not really relevant. A simple bound such as $|\rho_{nm}|<\Mreg$ for some regulator $\Mreg$ solves that problem by not allowing the coefficients to grow. Related to this, the dual problem is generically not feasible (feasibility depends on the functional chosen) unless we regulate the primal problem in which case the dual is feasible for any primal functional. Let us now consider these issues in more detail using the Mandelstam representation.

\subsection{Fredholm equations and regularization}

  A single or double dispersion relation defines the amplitude through a linear operator
\beq
 F = \cK \rho  \label{a30}
\eeq
Given an amplitude $F$, determining a spectral density $\rho$ is equivalent to solving a Fredholm equation of the first kind with kernel $\cK$ (see \eg\ \cite{groetsch1984theory}). For this problem to be well-posed we should require that for any $F$ there exists a unique $\rho$  and that the solution depends continuously on $F$. It is well-known \cite{groetsch1984theory}                                                                                                                                                                                                                                                                                                                                                                                                                                                                                                                                                                                                                                                                                                                                                    that this is not true when the kernel $\cK$ is continuous, the problem of finding $\rho$ is then ill-posed\footnote{A standard example of ill--posed problem is to solve the diffusion equation backwards in time.}. For example, adding a highly oscillating function to $\rho$, even with a large coefficient, does not modify $F$. In the case at hand, it means that the coefficients of such highly oscillating variations of $\rho$ cannot be determined by maximization since they barely modify the functional. On the other hand they also barely modify the partial waves and therefore have no physical interest and should be suppressed. 
For the case in hand, recall the definition of the amplitude \eqref{a2}. The single dispersion part has a singular kernel  
\beq
 \cK(s,x) \sim \frac{1}{x-s}, \ \ \ x\rightarrow s   \label{a31}
\eeq   
and therefore the problem is well-posed. In the double dispersion relations, however, if we look for example at the part of the kernel that behaves as
\begin{equation}
\mathcal{K}(s,t;x,y)\sim\frac{1}{(x-s)(y-t)},  
\label{a32}
\end{equation}
it is not singular for physical values of $t$ (\ie \ $4-s<t<0$) because in the region of integration in \eqref{a2} we have $y>4$ and then $y\neq t$. The same is true for other parts of the kernel in \eqref{a2}. 
 
 In the case of Fredholm equations of the first kind, the method to make the problem well-posed is to use the so-called Tikhonov regularization.  Instead of solving eq. \eqref{a30}, we minimize the functional
\beq
 \cF = ||\cK \rho-F||^2 + \gamma ||\rho||^2
\label{a33}
\eeq 
 where $||\cdot||$ is a norm\footnote{Note that the definition of the norm does not have to be the same for $\cK \rho-F$ and $\rho$.}. The parameter $\gamma$ provides a regularization. Ideally, the norm should be such that the variations of $\rho$ that affect the partial waves the most are kept and others are suppressed. In practice, the norms that we tried numerically, including the simple square norm, give similar answers. 
 
 In the following we discuss the dual problem that, if formulated in the usual way, leads to a linear constraint on the dual variables. Again, such constraint can be written as a Fredholm equation of the first kind. In 1+1 dimensions, such equation can be solved, but in 3+1 dimensions, the dual problem turns out to be weakly infeasible. The regularized case has no such problem. 
 
 As described in the next section, instead of the regularization in \eqref{a33}, it turns out that there is a similar form of regularization that fits perfectly well with convex maximization. 

\subsection{Regularized convex problem and its dual} \label{reg}
In this section we consider another motivation to regularize the primal problem. If we dualize a general conic optimization problem, the dual problem might not be feasible \cite{conic1}, \ie\ the dual constraints may have no solutions. In this case, regularizing the primal problem solves both issues, makes the dual problem feasible and suppresses unwanted variations of $\rho$ as discussed previously. Let us start with a generic discussion of regularization. 
  Consider real variables $v_j$ and the primal problem:
\begin{equation}
\begin{aligned}
\mmax{v_j}\;\;\Big\{\cF_P&= f_j v_j\Big\} \\ 
s.t.  \;\; A_{aj} v_j &=b_a,  \;\; \mathrm{and} \;\; 
v\in \cone
\label{a34}
\end{aligned}
\end{equation} 
 where $\cone$ is a cone. Note that by using extra variables $\eta_\ell(s)$ the unitarity constraint $|S_\ell(s)|\le 1$ can be written as a cone $|S_\ell(s)|^2 \le \eta_\ell(s)^2$ with a linear constraint $\eta_\ell(s)=1$. 
Writing the Lagrangian, we find that 
\beq
\begin{aligned}
L =&f_j v_j + y_a (A_{aj}v_j-b_a) + s_j v_j \\
=&(f_j+y_a A_{aj} +s_j) v_j -y_a b_a \\
\ge& f_j v_j\; , \ \ \ 
\label{a35}
\end{aligned}
\eeq
where $y_a$ are Lagrange multipliers and $s\in \cone^*$ where
\begin{equation}
\cone^*=\big\{s_j\;\big|s_jv_j\ge0,\forall v\in \cone\big\}
\label{a36}
\end{equation}
is the cone dual\footnote{We can also define $\cone^*$ without requiring that $\cone$ itself is a cone and everything works similarly.} to $\cone$. Thus, if we impose on $y_a$ and $s_j$ the constraints 
\begin{equation}
f_j+y_a A_{aj}+s_j=0,\; \;s\in \cone^*.
\label{a37}
\end{equation}
we find
\beq
\mmax{v_j}\; \Big\{\cF_P=f\cdot v\Big\} \le \mmin{y_a} \;\Big\{\cF_D=-b\cdot y\Big\}.
\label{a38}
\eeq 
 
However, in conic maximization, there is no guarantee that the dual problem is feasible, namely that the hyperplane $-f-y\cdot A$ parameterized by $y_a$ actually intersects $\cone^*$. For some matrices $A$ the hyperplane always intersects the cone, whereas for others it depends on the point $f_j$, namely on the primal functional. For a simple geometric picture one can imagine the future light-cone and lines in various directions and positions that may or may not intersect the cone. Once again, the solution to this issue is to regularize the primal problem. 

\medskip

Now we introduce two ways of regularizing that we call the $M-$ and $\gamma-$regularizations. In the $M-$regularization, we define the primal problem as
\begin{equation}
\begin{aligned}
\mmax{v_j}\; \Big\{\cF_P&=f_j v_j\Big\} \\
s.t. \;\; A_{aj}v_j &=b_a,\;\; v\in \cone,\\
||v||&\le \Mreg
\label{a39}
\end{aligned}
\end{equation}
for some norm $||\cdot||$ and a regulator $\Mreg$. We should introduce also the dual norm $||\cdot||_*$ defined as
\beq
||\mu||_* = \sup\big\{ \mu_j v_j\ \big|\  v\in \cone,\ ||v||\le 1\big\}
\label{a40}
\eeq
namely $\mu_j v_j \le ||v|| ||\mu||_*$ for any $v_j$, $\mu_j$. With this definition, and using $||v||\le \Mreg$ from \eqref{a39}, we can write
\begin{equation}
\begin{aligned}
L =&\ f_j v_j + y_a (A_{aj}v_j-b_a) + s_j v_j +||\mu||_* \Mreg - \mu_j v_j \\
\ge& f_j v_j ,
\label{a41}
\end{aligned}
\end{equation}
The dual constraints can now be solved as
\beq\label{regdual}
\mu_j = f_j+y_{a}A_{aj}+ s_j, 
\eeq
namely they just determine $\mu_j$. Replacing this result back in \eqref{a41} we get
\beq
\mmax{v_j}\;\Big\{\cF_P=f_j v_j\Big\}\le \mmin{y_a}\;\Big\{\cF_D=-y_a b_a + \Mreg ||f+y\cdot A+s||_*\Big\}
\label{a42}
\eeq

\smallskip

An alternative way of regularization, i.e., the $\gamma$-regularization is to consider
\begin{equation}
\begin{aligned}
\mmax{v_j} \;\Big\{ \cF_P&=f_j v_j - \gamma ||v|| \Big\}\ \\ 
s.t.  \;\;A_{aj} v_j &=b_a, \;\; v\in \cone 
\label{a43}
\end{aligned}
\end{equation}
If we now impose the conditions
\begin{equation}
\begin{aligned}
&\mu_j=f_j+y_aA_{aj}+s_j, \;\;s\in \cone^*\\
&||\mu_j||_*  \le \gamma
\end{aligned}
\label{a44}
\end{equation}
it follows that
\beqa
\cF_P &\le& f_j v_j -\gamma ||v|| + y_a (A_{aj}v_j-b_a) + s_j v_j +||\mu||_* ||v|| - \mu_j v_j \\
      &\le&  -y_a b_a + (f_j +y_aA_{aj} +s_j -\mu_j) v_j + (||\mu||_*-\gamma)\, ||v|| \\
      &\le& -y_a b_a
\label{a45}
\eeqa
namely
\beq
\mmax{v_j}\;\Big\{\cF_P=f_j v_j - \gamma ||v|| \Big\}\le \mmin{y_a}\;\Big\{\cF_D=-y_a  b_a \Big\}
\label{a46}
\eeq
subject to the constraints \eqref{a44}. The $M$- and $\gamma$- regularizations are actually dual of each other. In the following we choose the $M$-regularization but the $\gamma$-regularization is equally good. 

\subsection{The single pion case and its dual}\label{singlepi}

Now that we have discussed the idea of regularization and the primal problem we are going to construct the dual problem following the formal procedure known from standard convex maximization.
In the case of a single pion, the regularized primal problem is the one in \eqref{a24} with the extra condition $||\alpha||\le \Mreg$ for some norm $||\cdot||$. Thus, we want to solve
\begin{equation}
\begin{aligned}
\mmax{\alpha_n}  \;\Big\{ \cF_P&=a_n \alpha_n \Big\}\\
s.t.  \;\; h_\ell(s) &= \alpha_n h_{\ell n}(s),  \;\; \\
|S_\ell(s)| = |1+ih_\ell(s)|& \le 1, \;\;||\alpha||\le\Mreg 
\label{a47}
\end{aligned}
\end{equation} 
where $\ell$ is even.
In practice we increase the regulator $\Mreg$ until a plateau is reached in $\mmax{\alpha_n}\; \cF_P$ as a function of $\Mreg$. The norm $||\cdot||$ is largely arbitrary and can be chosen in different ways within reason as discussed below and in section \ref{sec5}.

\smallskip

Let us now write the Lagrangian
\beq
 L = \sum_n a_n \alpha_n + \intlevens \left[-\Re\big(\bar{k}_\ell(s) S_\ell(s)\big)+u_\ell(s)\right]
\label{a48}
\eeq
where the auxiliary functions $k_\ell(s)$ and $u_\ell(s)$ satisfy $|k_\ell(s)|\le u_\ell(s)$. 
It is then easy to see that
\beq
 L \ge \sum_n a_n \alpha_n = \cF_P
\label{a49}
\eeq
since
\begin{equation}
\text{Re}\big(\bar{k}_{\ell}(s)S_{\ell}(s)\big)\le|k_{\ell}(s)|\le u_{\ell}(s)
\end{equation}
where we have used the unitarity of $S_{\ell}(s)$.
Replacing $S_\ell(s)$ in terms of the partial waves we get
\beqa
 L &=& \intlevens \left[ u_\ell(s) -\Re k_\ell(s)\right] +\langle \alpha, \Theta\rangle  \label{a50}\\
 \Theta_n &=& a_n + \intlevens \Im\left(\bar{k}_\ell(s) h_{\ell,n}(s)\right) \label{a51}
\eeqa
where we defined the scalar product
\beq
\langle \alpha, \Theta\rangle = \sum_n \alpha_n \Theta_n \label{a52}
\eeq
 Given the constraint $||\alpha||\le \Mreg$, and from the definition of dual norm \eqref{a40}, we have 
 \begin{equation}
 \langle \alpha, \Theta\rangle\le ||\alpha||\;||\Theta||_*\le \Mreg ||\Theta||_* \ ,
 \end{equation}
which, together with the condition 
\beq \label{a52b}
|k_\ell(s)|\le u_\ell(s)\, ,
\eeq
 leads to
\beq
\begin{aligned}
	 \cF_P = \sum_n a_n\alpha_n \le \intlevens \left[|k_\ell(s)|-\Re k_\ell(s)\right] + \Mreg||\Theta||_* = \cF_D \ .  
	\label{a53}
\end{aligned}
\eeq
where we chose the lowest allowed value of $u_\ell(s)$, \ie\ $u_\ell(s)=|k_\ell(s)|$.
 Thus the dual problem is 
\beq
\begin{aligned}
	 \mmin{k_\ell(s)}\;\; \bigg\{ \cF_D &= \intlevens \left[|k_\ell(s)|-\Re k_\ell(s)\right] + \Mreg||\Theta||_* \bigg\}\\
	 s.t.\;\;\Theta_n &= a_n + \intlevens \Im\left(\bar{k}_\ell(s) h_{\ell,n}(s)\right) \ .
	\label{a56}
\end{aligned}
\eeq 
Although replacing  $u_\ell(s)=|k_\ell(s)|$ makes the expression more compact, in order to use standard optimization software for the numerical part it is better to keep that part of the objective linear as in \eqref{a50} and require \eqref{a52b}. The same can be done with $||\Theta||_*$ by introducing a variable $\nu$ and requiring $||\Theta||_*\le\nu$, \ie\ the equivalent linear objective used for the numerical part reads 
\beq
\cF_D = \sum_{\ell \mathrm{even}} \int_4^\infty\! ds\, \left[u_\ell(s)-\Re k_\ell(s)\right] + \Mreg\, \nu
\eeq
with constraints $|k_\ell(s)|\le u_\ell(s)$, and $||\Theta||_*\le\nu$. Also, as in the primal problem \eqref{a47}, to choose a regulator $\Mreg$, we increase its value until we reach a plateau.
 The $\Theta_n$ associated with $f_0$ and $\sigma(x)$ can be set to zero since that part of the problem is feasible: 
\begin{subequations}
\beqa
 \Theta_0 &=& a_0 + \frac{\pi}{2} \int_4^\infty\!\!\!\!\!\! ds\, \sqrt{\frac{s-4}{s}}\, \Im \bar{k}_0(s) =0  \label{a58} \\
 \Theta(x) &=& a(x)+\intlevens \Im\left(\bar{k}_{\ell}(s) h_\ell(x)\right)  =0 \ , \label{a59}
\eeqa
\end{subequations} 
 which can be seen as a choice of norm for those components. As discussed later in \eqref{a116}, this cannot be done for the double dispersion relation part
\beq
 \Theta(x,y)= a(x,y) +\intlevens \Im\left(\bar{k}_{\ell}(s) h_\ell(x,y) \right) \ .
\label{a57}
\eeq  
 For those components a natural dual norm is given by
 \beq
  ||\Theta||_* =\intlevens\  \left| \intxy \Im h_{\ell}(x,y;s) \Theta(x,y) \right| \ ,
 \label{a54}
 \eeq
 that suppresses the $\Theta(x,y)$ associated with variations that affect the most the imaginary part of the partial waves. For practical purposes we can choose other norms since the results should be independent of the specific regularization. 
 Since this is a bounded convex problem and, after regularizing, the dual is feasible, the duality gap closes \cite{conic1}\footnote{In the continuum case, like here, the closing of the duality gap can be seen as a consequence of the Hahn-Banach theorem \cite{duren1970}. This also applies to the numerical discretization as also follows from general arguments for conic optimization \cite{conic1}.}, namely
\beq
 \mmax{\alpha_n}\, \cF_P = \mmin{k_\ell(s)}\, \cF_D 
\eeq
 In that case we have $-\Re(\bar{k}_\ell(s) S_\ell(s))+|k_\ell(s)|=0$. Thus 
\beq
  S_\ell(s) = \frac{k_\ell(s)}{|k_\ell(s)|}, \ \ \ \Rightarrow \ \ \ h_\ell(s)=-i\left(\frac{k_\ell(s)}{|k_\ell(s)|}-1\right)
\label{a55}
\eeq
Since the $S_\ell(s)$ are the partial waves of the primal problem, they automatically satisfy crossing and analyticity when the minimum is found exactly. On the other hand, as long as the relatively mild constraints \eqref{a58}, \eqref{a59} are satisfied, the $k_\ell(s)$ can be chosen arbitrarily and therefore, specially in the numerical procedure, we can truncate the space of $k_\ell(s)$ to a finite number of partial waves $\ell$ and discretized energy $s_j$. In this case the minimum has a gap with the maximum of the primal problem and therefore analyticity and crossing are satisfied only approximately. Notice that taking $k_\ell(s)=0$ implies that the corresponding $S_\ell(s)$ is undefined. Finally, if the duality gap closes we have $\langle\alpha,\Theta\rangle=\Mreg ||\Theta||_*$ which allows to compute the primal variables $\alpha_n$.

\subsection{The $O(N)$ case and its dual}\label{ONd}
The $O(N)$ case is a simple generalization of the previous case. We obtain for the primal problem
\begin{equation}
\begin{aligned}
\mmax{\alpha_n} \;\; \cF_P&= n_I \cF^I(s_0,t_0,u_0) = a_n \alpha_n, \\ 
s.t.  \;\; h^I_\ell(s)& = \alpha_n h^I_{\ell n}(s) \\ \;\; |S^I_\ell(s)| = |1+ih^I_\ell(s)| &\le 1\;\;,||\alpha|| \le \Mreg
\end{aligned}
\label{a62}
\end{equation} 
and for the dual
\begin{equation}
\begin{aligned}
\mmin{k^I_\ell(s)} &\;\;\bigg\{ \cF_D=\intlsI \left[|k^I_\ell(s)|-\Re k^I_\ell(s)\right] + \Mreg\, ||\Theta||_* \bigg\} \\
s.t. & \;\; \Theta_n = a_n + \intlsI \Im\left(\bar{k}^I_\ell(s) h^I_{\ell,n}(s)\right)
\end{aligned}
\label{a63}
\end{equation} 
Again we can set to zero the $\Theta$ associated with $f_0$ and $\sigma_\alpha(x)$:
\begin{subequations}
\beqa
\Theta_0 &=& (N+2)n_0 + 2 n_2 + \frac{\pi}{2} \int_4^\infty\!\!\!\!\!\! ds\, \sqrt{\frac{s-4}{s}}\, \Im\left[(N+2)\bar{k}^{I=0}_0(s)+2\bar{k}^{I=2}_0(s) \right] =0 \non \label{b58} \\
\Theta_\alpha(x) &=& a_\alpha(x)+\intlsI \Im\left(\bar{k}^I_{\ell}(s) h^I_{\ell,\alpha}(s)\right)  =0  \label{b59}
\eeqa
\end{subequations} 
and choose a norm for the double dispersion relation part associated with $\rho_\alpha(x,y)$:
\beq
\Theta_\alpha(x,y)= a_\alpha(x,y) +\intlsI \Im\left(\bar{k}^I_{\ell}(s) h^I_{\ell,\alpha}(x,y;s) \right)
\label{b57}
\eeq  
In the previous expressions the sums are over $I=0,1,2$ and over $\ell$ even for $I=0,2$ and $\ell$ odd for $I=1$. The partial waves can be computed as in \eqref{a55}:
\beq
h^I_\ell(s)=-i\left(\frac{k^I_\ell(s)}{|k^I_\ell(s)|}-1\right)
\label{c55}
\eeq 
Once again they lead to partial waves that saturate unitarity unless $k^I_\ell(s)$ vanishes for some values of $s$ in which case the partial waves are undetermined for those $s$ values.

\section{Generalized dispersion relations and the dual amplitudes}
\label{sec4}

In 1+1 d, an important part of the dual problem is that it can be derived using generalized dispersion relations \cite{pollica}. The main idea is that the functional  is the value of the analytic amplitude in the unphysical kinematic region which, by using dispersion relations, can be written in terms of the boundary values, namely, the values in the physical region that are known to obey the unitarity constraints. In 3+1 d, when using double dispersion relations, we are not aware of a similar construction. Therefore, in this section, we derive a generalized double dispersion relation \eqref{formula} which can be used to relate the value of the amplitude at an unphysical point to the values on the physical region, plus an extra term which can be bounded and can be made as small as required by appropriately choosing the function in the dispersion relation.

\subsection{Dual problem in 1+1 d from dispersion relations}
Suppose we have an S-matrix $S(s)$ given by an analytic function with cuts for $s>4$ and $s<0$. We propose the dispersion relation
\beq
S(s) = \frac{1}{\pi} \intx \frac{\rho(x)}{x-s} + \frac{1}{\pi} \intx \frac{\rho(x) }{x-4+s}
\label{singledisp}
\eeq
satisfying crossing: $S(s)=S(4-s)$. Consider maximizing a functional $\cF_P=  \Re S(s_0)$ with  $0<s_0<4$ under the unitarity constraint $|S(s^+)|\le 1$ for all $s>4$. 
We can write 
\beq
 \cF_P =  \Re S(s_0) = \Re\left[ \frac{1}{2\pi i}\oint_{\cC} \frac{S(z)}{z-s_0} dz\ \right]
\label{a65}
\eeq
where $\cC$ is a small contour encircling $s_0$. By deforming the contour we obtain the dispersion relation \eqref{singledisp}.
The observation is that we can replace $\frac{1}{z-s_0}$
with any analytic function $K(z)$, as long as it has a simple pole at $s_0$ with residue one, and no other singularities except possible cuts on the real axis at $s<0$ and $s>4$. Then we can write
\beq
\cF_P =  \Re  S(s_0) = \Re\left[ \frac{1}{2\pi i}\oint_{\cC} S(z)K(z) dz\ \right]
\label{a66}
\eeq
 For simplicity we assume (anti)-crossing $K(s)=-K(4-s)$ which requires adding a pole at $4-s_0$, real analyticity $K^*(s)=K(s^*)$, and that $K(s)$ falls sufficiently fast at infinity so that we can deform the contours $\cC$ to wrap the cuts. After some algebra we find
\beq
\begin{aligned}
	 \cF_P =& \frac{2}{\pi} \int_4^\infty  \Im\left[ S(s) K(s) \right] ds\\
	\le& \frac{2}{\pi} \int_4^\infty \left| S(s^+) K(s^+) \right| ds\\
	\le& \frac{2}{\pi} \int_4^\infty | K(s^+) | ds \\
	=&\cF_D
	\label{a67}
\end{aligned}
\eeq 
 The best bound is obtained from minimizing $\cF_D$ over all possible functions $K(z)$. 

\subsection{Generalized dispersion relations in 1+1 d}

The same result can be obtained by using a similar but slightly different formulation of the problem that generalizes directly to 3+1 dimensions. 

We start by considering analytic functions defined through a dispersion relation\footnote{In our applications $g(x)$ only has support on part of the real axis but here this property is not necessary.}  
\beq
 G(z) = \frac{1}{\pi} \int_{-\infty}^{+\infty}\!\!\! dx \frac{g(x)}{x-z} 
\label{Gdisp}
\eeq
such that 
\begin{equation}
\Delta G(x) \equiv G(x^+)-G(x^-) = 2i g(x).
\label{gdisc}
\end{equation}
 Regarding crossing symmetry, notice that $g(4-x)=\pm g(x)$ implies $G(4-z) = \mp G(z)$. 
 For any such function we can use contour integration assuming that we can drop the contribution from the arc at infinity to derive
\begin{subequations}\label{identities}
\begin{eqnarray}
 \int_{-\infty}^{+\infty}\!\!\! dx\ G(x^+) &=& 0 \\
 \int_{-\infty}^{+\infty}\!\!\! dx\ G(x^-) &=& 0
\label{a70}
\end{eqnarray}
\end{subequations}

Take now two such functions: $H(z)$ given by
\begin{equation}
H(z) = \frac{1}{\pi} \int_{-\infty}^{+\infty}\!\!\! dx \frac{h(x)}{x-z} 
\label{a71}
\end{equation}
and $G(z)$ as written in \eqref{Gdisp}. We have
\begin{equation}
\begin{aligned}
 &\int_{-\infty}^{+\infty}\!\!\! dx\ \left[\Delta G(x) H(x^+) +\Delta H(x) G(x^-)\right]\\
= &\int_{-\infty}^{+\infty}\!\!\! dx\ \left[\big(G(x^+)-G(x^-)\big) H(x^+) + \big(H(x^+)-H(x^-)\big) G(x^-) \right]\\
= & \int_{-\infty}^{+\infty}\!\!\! dx\ \left[ G(x^+) H(x^+) - H(x^-) G(x^-)\right]\\
 =& 0
\end{aligned}
\label{a72}
\end{equation}
where the last identity follows from the fact that $G(z)H(z)$ is analytic in the upper half-plane and also in the lower half-plane as was the case in (\ref{identities}). 
Using 
\begin{equation}
\Delta H(x) \equiv H(x^+)-H(x^-) = 2i h(x).
\label{a73}
\end{equation}
and \eqref{gdisc}, we then find
\beq
  \int_{-\infty}^{+\infty}\!\!\! dx\ \left[g(x) H(x^+) + h(x) G(x^-)\right] =0
\label{a74}
\eeq

Now consider a crossing symmetric analytic function $S(z)$ whose discontinuity is given by
\beq
h(x) = -\theta(-x) \rho(4-x) + \theta(x-4) \rho(x) 
\label{a75}
\eeq
where $\theta$ is the Heaviside step function, and for $g(z)$ we take
\beq
 g(x) = - \pi \delta(x-s_0) - \pi \delta(x-4+s_0) + \theta(-x) k^*(4-x) + \theta(x-4) k(x) 
\label{a76}
\eeq
such that $g(4-x)=g^*(x)$ or $G^*(4-z)=-G(z^*)$. 
 In this case \eqref{a74} reduces to	
\beq
\begin{aligned}
	 \cF_P =& \Re S(s_0)\\
	  =& \frac{1}{\pi} \int_4^\infty\!\!\! dx\, \Re\left[k(x)S(x^+)\right] + \frac{1}{\pi} \int_4^\infty\!\!\! dx\, \rho(x) \Re\, G(x^-)  
	\label{a77}
\end{aligned}
\eeq
 In 1+1 d we have two options. We can regularize the problem by requiring $||\rho||<\Mreg$ and then obtain the bound
\beq
 \cF_P \le \frac{1}{\pi} \int_4^\infty dx |k(x)| + \frac{\Mreg}{\pi} ||\Theta(x)||_*
\label{a78}
\eeq 
where $\Theta(x)=\Re \,G(x^-)$. In this case the dual functional contains $|k(x)|$ which is not the boundary value of an analytic function, but a spectral density as in \eqref{a76}. Another possibility is to impose $\Re\, G(x^-)=0$ for $x>4$ instead. In this case, using \eqref{gdisc} and \eqref{a76}, one can compute the imaginary part of $k(x)$ in terms of its real part. The result is
\beqa
 k(x) &=& -\frac{i}{\pi}\int_{-\infty}^{\infty} \frac{\kappa(\xi)}{\xi-x^+}d\xi \\
\kappa(\xi) &=& -\pi\delta(\xi-s_0)-\pi\delta(\xi-4+s_0)\label{a79}\\
&&+\theta(-\xi)\Re k(4-\xi)+\theta(\xi-4) \Re k(\xi)\nonumber
\eeqa
Since $\kappa(\xi)\in\mathbb{R}$ and $\kappa(4-\xi)=\kappa(\xi)$, the function 
\beq
 \tilde{K}(z) = \frac{1}{\pi} \int_{-\infty}^{\infty} \frac{\kappa(\xi)}{\xi-z}d\xi 
\label{a80}
\eeq
defines what we call the dual amplitude. It is anti-crossing-symmetric, real analytic, has poles with residue one at $s_0$ and $4-s_0$ and cuts on $(-\infty,0)\cup (4,\infty)$. Its boundary value above the cut $(4,\infty)$ is $k(x)$ from where we can compute the S-matrix. Therefore we go back to the previous case. There is however an important difference between the two cases: in the first case, $k(x)$ is not the boundary value of an analytic function but its jump across the cut. Therefore, if it vanishes on a segment, it does not necessarily vanish everywhere. This is important since vanishing of $k(x)$ allows unitarity of $S(x)$ in the same region not to be saturated. Therefore imposing a restriction on $\rho(x)$ (such as $||\rho||\le \Mreg$) allows the duality gap to close without implying unitarity saturation at all energies.

\subsection{Generalized dispersion relations in 3+1 d and the dual amplitude}\label{gdr}

 In the case of 3+1 dimensions we have amplitudes that depend analytically on two independent Mandelstam variables $(s,t)$. We assume they obey double dispersion relations with support on the real axis of $s$ and $t$. We are going to ignore subtractions, since the single dispersion relations can be treated as the 1+1 d case. Let us consider analytic functions of the form
\beq
G(w_1,w_2) = \frac{1}{\pi^2} \intxyi \frac{g(x,y)}{(x-w_1)(y-w_2)}
\label{a81}
\eeq
and define the double jump for $x,y\in \mathbb{R}$:
\beq
 \Delta_{12} G(x,y) = G(x^+,y^+)-G(x^-,y^+)-G(x^+,y^-)+G(x^-,y^-) = -4 g(x,y)
\label{dd}
\eeq
 For any function such as $G(w_1,w_2)$, analytic in the upper and lower half-planes (excluding the real axis) and assuming we can drop the arc at infinity, we get
 \begin{subequations}
 	\beqa
 	\int_{-\infty}^{+\infty}\!\!\!\!\!\! dx\ G(x^\pm,w_2) &=& 0 \label{b83} \\
 	\int_{-\infty}^{+\infty}\!\!\!\!\!\! dy\ G(w_1,y^\pm) &=& 0 \label{a83}
 	\eeqa
 \end{subequations}
Multiplying any two such functions $H,K$ and using \eqref{dd}, we find
\beq 
 \intxyi \left[\Delta_{12} H(x,y) K(x^-,y^+) - H(x^+,y^-) \Delta_{12} K(x,y) \right] = 0
\label{formula}
\eeq
where we used \eqref{b83}, \eqref{a83} to drop all terms where the integral over either $x$ or $y$ vanished. We call \eqref{formula} a generalized dispersion relation since, as we will see below, it can be used to relate the values of the analytic amplitude at an unphysical point to the values in the physical region.   

\smallskip

Using the identity \eqref{formula}, we can now extract from the analytic amplitude its value on the physical region $s>4, 4-s<t<0$, by introducing an analytic function, \ie, the 3+1 d dual amplitude
\beq\label{Kdisp}
K^{st}(z,w) = -\frac{1}{(z-s_0)(w-t_0)} + \frac{i}{\pi^2} \intst \frac{\bar{k}(s,t)} {(z-s)(w-t)}
\eeq
so that\footnote{With a slight abuse of notation we define the function $\theta$ to be one when all conditions in its argument are met and zero otherwise.}
\beq
\Delta_{12} K^{st}(x,y) = 4\pi^2\delta(x-s_0)\delta(y-t_0) - 4i\bar{k}(x,y) \theta(x>4,4-x<y<0)
\label{a85}
\eeq
It is important to note that we do not take $\bar{k}(x,y)$ to be real. Now, take an amplitude $H^{st}(s,t)$ to be given by
\beq
 H^{st}(s,t) = \frac{1}{\pi^2} \intxyi \frac{\rho^{st}(x,y)}{(x-s)(y-t)}
\label{a86}
\eeq
so that $\Delta_{12} H^{st} =-4 \rho^{st}$, and assume that $\rho^{st}(x,y)$ vanishes\footnote{This is used only to replace $H(s^+,t^-)=H(s^+,t)$ in the physical region.}  in the physical region $x>4, 4-x<y<0$. Then \eqref{formula} gives
\beq
H^{st}(s_0,t_0) = \frac{i}{\pi^2}\intst H^{st}(s^+,t) \bar{k}(s,t) - \frac{1}{\pi^2} \intxyi \rho^{st}(x,y) K^{st}(x^-,y^+) 
\label{a87}
\eeq
In the physical region $s>4,\,4-s<t<0$, the values $H^{st}(s^+,t)$ can be written in terms of its partial waves as in \eqref{Hinv}:
\beq
 H^{st}(s^+,t) =  \frac{2}{\pi}\sqrt{\frac{s}{s-4}} \sum_{\ell } (2\ell+1) h^{st}_\ell(s) P_\ell\left(1+\frac{2t}{s-4}\right)
\label{a88}
\eeq 
resulting in 
\beq
H^{st}(s_0,t_0) = i \intls h^{st}_\ell(s) \bar{k}_\ell(s) -\frac{1}{\pi^2} \intxyi \rho^{st}(x,y) K^{st}(x^-,y^+) 
\label{a89}
\eeq
where we have defined the dual partial waves $k_\ell(s)$ as
\beq
k_\ell(s) = \frac{(2\ell+1)}{\pi^3}\sqrt{s(s-4)}  \intP k(s,t)
\label{a90}
\eeq
and as usual, $t$ in the integrand is a function of $\mu,s$ as in \eqref{a4}. We can invert this last relation as
\beq
 k(s,t)=\frac{\pi^3}{2\sqrt{s(s-4)}}\sum_\ell k_\ell(s) P_\ell\left(1+\frac{2t}{s-4}\right)
\label{a91}
\eeq
When the minimum of the dual problem is achieved, the dual partial waves are directly related to the physical partial waves as seen in \eqref{a55}. 
Taking the real part in \eqref{a89} we find
\beq
 \Re H^{st}(s_0,t_0) = -\intls\Im[h_\ell^{st}(s) \bar{k}_\ell(s)] +\intxyi \rho^{st}(x,y) \Theta^{st}(x,y) 
\label{a92}
\eeq
with 
\beq
 \Theta^{st}(x,y) = -\frac{1}{\pi^2}\Re K^{st}(x^-,y^+)
\label{a93}
\eeq
For comparison with \eqref{a57} it is convenient to define
\beqa
h^{st}_\ell(x,y;s) &=& \frac{1}{4\pi}\sqrt{\frac{s-4}{s}}\frac{1}{x-s^+}\int_{-1}^{+1}\frac{P_{\ell}(\mu)}{y-t} \nonumber\\
                   &=& \frac{1}{\pi} \frac{1}{\sqrt{s(s-4)}} \frac{1}{x-s^+} \Qy
\label{a94}
\eeqa 
namely the coefficient of the partial waves:
\beq
 h^{st}_\ell(s) = \intxyi\ \rho^{st}(x,y) h_\ell^{st}(x,y;s)
\label{a95}
\eeq
With these coefficients we can write
\beq
\frac{1}{\pi^2}\frac{1}{(x-s^+)(y-t)}=\frac{2}{\pi}\sqrt{\frac{s}{s-4}} \sum_\ell (2\ell+1) h^{st}_\ell(x,y;s) P_\ell\left(1+\frac{2t}{s-4}\right)
\label{a96}
\eeq
Replacing in \eqref{Kdisp} and after some algebra we obtain from \eqref{a93}
\beq
 \Theta^{st}(x,y) = \frac{1}{\pi^2}\frac{1}{(x-s_0)(y-t_0)} +  \intls \Im[\bar{k}_\ell(s) h^{st}_\ell(x,y;s)] 
\label{a97}
\eeq
which is the same as formula \eqref{a57}. 

\smallskip

Suppose now that, we have an amplitude defined by
\beq
H^{su}(s,u) = \frac{1}{\pi^2} \intxyi \frac{\rho^{su}(x,y)}{(x-s)(y-u)}
\label{b86}
\eeq
and the dual amplitude
\beq
 K^{su}(z,w) = -\frac{1}{(z-s_0)(w-u_0)} + \frac{i}{\pi^2} \intst \frac{\bar{k}(s,t)} {(z-s)(w-u)}
\label{a98}
\eeq
where $\bar{k}(s,t)$ is the same function as in \eqref{Kdisp} and $u$ in the integrand should be understood as $u=4-s-t$. The same procedure leads to
\beq
\Re H^{su}(s_0,u_0) = -\intls\Im[h_\ell^{su}(s) \bar{k}_\ell(s)] + \intxyi \rho^{su}(x,y) \Theta^{su}(x,y) 
\label{a99}
\eeq  
with 
\beq
\Theta^{su}(x,y) = \frac{1}{\pi^2}\frac{1}{(x-s_0)(y-u_0)} +  \intls \Im[\bar{k}_\ell(s) h^{su}_\ell(x,y;s)] 
\label{a100}
\eeq
where the partial waves are defined in the usual manner \eqref{hldef} and \eqref{a6}. Thus, for an amplitude of the form
\beq
 F(s,t,u) = \frac{1}{\pi^2} \intxyi \left[\frac{\rho^{st}(x,y)}{(x-s)(y-t)} + \frac{\rho^{su}(x,y)}{(x-s)(y-u)}\right]
\label{a101}
\eeq
we find
\beqa
\Re F(s_0,t_0,u_0) &=& -\intls\Im[h_\ell(s) \bar{k}_\ell(s)]  \label{b1}\\
                   && + \intxyi \left[\rho^{st}(x,y) \Theta^{st}(x,y) + \rho^{su}(x,y) \Theta^{su}(x,y)\right] \non
\label{a102}
\eeqa

\medskip

Now we are ready to consider an amplitude of the generic Mandelstam form
\beq
 F(s,t,u) = \frac{1}{\pi^2} \intxy \left[\frac{\rho_1(x,y)}{(x-s)(y-t)}+\frac{\rho_2(x,y)}{(x-s)(y-u)}+\frac{\rho_3(x,y)}{(x-t)(y-u)}\right] 
\label{a103}
\eeq
Using the identity
\beq
\frac{1}{(x-t)(y-u)} = \frac{1}{(x+y-4+s)(x-t)}+\frac{1}{(x+y-4+s)(y-u)} 
\label{a104}
\eeq 
eq. \eqref{a103} can be written in the form of \eqref{a101} with
	\beq
	\begin{aligned}
			\rho^{st}(x,y) &= \rho_1(x,y)\theta(x>4,y>4) - \rho_3(y,4-x-y) \theta(y>4,x+y<0)  \\
	    	\rho^{su}(x,y) &= \rho_2(x,y)\theta(x>4,y>4) - \rho_3(4-x-y,y) \theta(y>4,x+y<0) 	 \label{a106}
	\end{aligned}
	\eeq
We then have
\beq
\Re F(s_0,t_0,u_0) = -\intls\Im[h_\ell(s) \bar{k}_\ell(s)] + \sum_{a=1}^{3} \intxy \rho_a(x,y) \Theta_a(x,y)
\label{a107}
\eeq
with
\begin{subequations}
	\beqa
	\Theta_1(x,y) &=&  \Theta^{st}(x,y)  \label{a108}\\
	\Theta_2(x,y) &=&  \Theta^{su}(x,y)  \label{a109}\\
	\Theta_3(x,y) &=& -\Theta^{st}(4-x-y,x)-\Theta^{su}(4-x-y,y) = -\frac{1}{\pi^2}\Re K^{tu}(x^-,y^+) \label{a110}\nonumber\\
	\eeqa
\end{subequations}
where $K^{tu}$ is given by
\beqa
 K^{tu}(z,w) &=& -K^{st}(4-z-w,z)-K^{su}(4-z-w,w) \non \\
             &=& -\frac{1}{(z-t_0)(w-u_0)} + \frac{i}{\pi^2} \intst \frac{\bar{k}(s,t)} {(z-t)(w-u)}
 \label{b98}
\eeqa
Defining a scalar product as 
\beq
 \langle \rho\cdot\Theta\rangle = \intxy \rho(x,y)\Theta(x,y) 
\eeq
 and using the chain of inequalities
\beq
 -\Im[h_\ell(s) \bar{k}_\ell(s)] = \Re[S_\ell k_\ell-k_\ell] \le |S_\ell k_\ell| -\Re k_\ell \le |k_\ell| -\Re k_\ell 
 \label{a112}
\eeq
we get the bound for \eqref{a107}:
\beq
 \Re F(s_0,t_0,u_0) \le \intls \left[|k_\ell| -\Re k_\ell\right] + \sum_{a=1}^{3}\langle \rho_a,\Theta_a\rangle
\label{a113}
\eeq 

\bigskip

At this point we might want to set $\Theta_a(x,y)=0$ for $a=1,2,3$ and $x>4$, $y>4$ so that we remove the $\rho_a$'s from eq. \eqref{a113}. This is however not possible. By interchanging the order of integration in  \eqref{Kdisp} one can write
\beqa
\Theta_1(x,y)  &=& \frac{1}{\pi^2} \frac{1}{(x-s_0)(y-t_0)} -  \frac{1}{\pi^2} \int_{-\infty}^0\!\!\!\!\!\! dt \frac{f(x,t)}{y-t} \label{a114}\\
f(x,t) &=&   \int_{4-t}^\infty\!\!\!\!\!\! ds\, \Im\left[ \frac{\bar{k}(s,t)} {x-s^+}  \right]
\label{a115}
\eeqa
 Setting $\Theta_1(x,y)=0$ implies, for every $y>4$ 
\beq
 \int_{-\infty}^0\!\!\!\!\!\! dt \frac{(x-s_0) f(x,t)}{y-t} = \frac{1}{y-t_0} 
\label{a116}
\eeq 
 The left hand side defines an analytic function of $y$ with a cut on the $(-\infty,0]$ line. If such function equals the right hand side for $y>4$, then, by analytic continuation they are equal everywhere. However the left hand side is finite when $y=t_0$ and therefore has no pole at $y=t_0$. Thus the equality is impossible and the dual problem where $\Theta_1=0$ is infeasible. 
 To elaborate this further, let us map the plane with a cut on the $(-\infty,0]$ line to the unit disk using
\beq
 w= \frac{2-\sqrt{y}}{2+\sqrt{y}}
\label{w1}
\eeq
 The region $y\ge4$ maps to the segment $[-1,0]$ and $t_0$ to a point $w_0$ on the positive real axis inside the disk. The function $\frac{1}{y-t_0}$ on the right-hand side of \eqref{a116} maps to the continuous function $\frac{1}{y(w)-t_0}$ for $w\in[-1,0]$ and therefore can be approximated arbitrarily close by a polynomial of high enough degree. Since a polynomial is an analytic function, when mapped back to $y$, we can represent it as a Cauchy integral and thus write it as in the left-hand side of \eqref{a116}. That means that we can make both sides of \eqref{a116} arbitrarily close. However, once again, if we want to make them exactly equal we need a polynomial of infinite degree, namely a series expansion. We know that such series expansion has to diverge at $w_0$ and does not define an analytic function inside the disk. This means that the dual problem is weakly infeasible. It is also true that for some particular primal functionals, namely some particular right-hand side of equation \eqref{a116}, the problem can be solved and in that case the dual is feasible. One simple example is a functional written in terms of a few partial waves as in \eqref{a19}. In that case, if we choose $\bar{k}^I_\ell(s)=-ic^I_\ell(s)$ in \eqref{b57} then $\Theta_\alpha(x,y)=0$. Furthermore, in that case, the functional $\cF_D$ is finite since only a finite number of the $c^I_\ell(s)$ are non-vanishing. This is an important distinction, since the functional \eqref{a7} can be written as in \eqref{a19} but with an infinite number of coefficients as explained in appendix \ref{appendix}. On the other hand, given that regularizing the primal problem is also important from the perspective of section \ref{sec3}, we are going to always consider regularized primal problems.

\subsection{The dual amplitude for single pion}

For the single pion all $\rho_a(x,y)$ are the same and symmetric. Therefore we get
\beqa
 \cF_P &=& \Re F(s_0,t_0,u_0) \le \intlevens \left[|k_\ell| -\Re k_\ell\right] + \langle \rho,\Theta\rangle \label{a117}\nonumber\\
       &\le&  \intlevens \left[|k_\ell| -\Re k_\ell\right] +  \Mreg ||\Theta||_* =\cF_D
       \label{a118}
\eeqa
with 
\beqa
\Theta(x,y)&=&\half \big(\Theta_1(x,y)+\Theta_2(x,y)+\Theta_3(x,y)\big) +(x\leftrightarrow y) \nonumber\\
           &=& -\frac{1}{\pi^2}\Re K(x^-,y^+)
\label{a119}
\eeqa
and $\Theta_{1,2,3}(x,y)$ are those in \eqref{a108}-\eqref{a110}. We also assumed $||\rho||\le \Mreg$ and $|S_\ell(s)|\le 1$. The dual amplitude is
\beqa
 K(z,w) &=& \half \left[ K^{st}(z,w)+K^{su}(z,w)+K^{tu}(z,w)\right] + (z\leftrightarrow w) \label{Kdualsp}\\
        &=& -\pi^2\,\cK(s_0,t_0,u_0;z,w) + i \intst \bar{k}(s,t) \cK(s,t,u;z,w) \non
\eeqa
where we used the kernel \eqref{a3b}. Replacing $\Theta_a(x,y)$ in \eqref{a119} one can see that $\Theta(x,y)$ is the same as the one in \eqref{a57} and therefore only contains $k_\ell(s)$ with $\ell$ even. The same is true for the first term since in \eqref{a107}, for this case, $h_\ell(s)=0$ if $\ell$ is odd. Thus, we rederive the dual problem using the generalized dispersion relations.  Although the result is the same as with the standard conic optimization method in section \ref{singlepi}, the main result of this section is the existence of the dual amplitude \eqref{Kdualsp}, an analytic function of two variables that also captures all information on the physical partial waves. Namely, from its double jump $k(s,t)$ one can compute the partial waves $f_\ell(s)$ through eqs.(\ref{a91}) and (\ref{a55}). Thus, it contains the same physical information as $F(s,t,u)$. A caveat is that the partial waves extracted this way from \eqref{a55} always saturate unitarity, unless the dual partial waves $k_\ell(s)$ vanish for some range of $s$. In such case the partial waves cannot be obtained from the dual amplitude and the corresponding range of energy allows the possibility of unitarity unsaturation. We leave more explorations in this direction for future work.

\subsection{The dual amplitudes for the $O(N)$ model}\label{ONcomplex}
For the $O(N)$ model the primal functional is
\beq
 \cF_P = n_I F^I(s_0,t_0,u_0) 
\label{a120}
\eeq
Each $F^I$ can be treated independently, therefore we are led to the dual amplitudes
\beq
	K^{st,I}(z,w) = -\frac{n^I}{(z-s_0)(w-t_0)} + \frac{i}{\pi^2} \intst \frac{\bar{k}^I(s,t)} {(z-s)(w-t)} \label{a121}\\
\eeq
and the same for $K^{su,I}(x,y)$ and $K^{tu,I}(x,y)$. Then
\beq
  \cF_P = n_I F^I(s_0,t_0,u_0) = -\intlsI \Im[h^I_\ell(s) \bar{k}^I_\ell(s)] + \sum_{I=0}^2\sum_{a=1}^{3} \langle \rho^I_a, \Theta^I_a \rangle  
\label{a123}
\eeq
 In this problem there are only two independent densities: $\rho_1(x,y)$, and $\rho_2(x,y)=\rho_2(y,x)$. This allows us to write
\beqa
 \cF_P &\le& \intlsI [|k^I_\ell(s)| -\Re k^I_\ell(s)] + \langle\rho_1, \Theta_1 \rangle + \langle\rho_2, \Theta_2 \rangle \label{a124}\nonumber\\
       &\le& \intlsI [|k^I_\ell(s)| -\Re k^I_\ell(s)] + \Mreg (||\Theta_1||_* + || \Theta_2||_*) \label{a125}
\eeqa
the functions $\Theta_{1,2}(x,y)$ are linear combinations of the $\Theta^I_a(x,y)$ such that this expression reduces to \eqref{a63}. More explicitly
\beq
 \Theta_\alpha(x,y) = -\frac{1}{\pi^2} \Re K_\alpha(x^+,y^-)
\eeq
with the dual amplitudes ($\alpha=1,2$)
\beqa
K_\alpha(z,w) &=& -\pi^2\, n_I\, \cK^I_\alpha(s_0,t_0,u_0;z,w)\nonumber \\
 && + i \intst\ \bar{k}^I(s,t)\, \cK^I_\alpha(s,t,u;z,w) 
\label{KdualON}
\eeqa
with an implicit sum over $I$ and the kernels $\cK_\alpha^I$ are those in \eqref{c1}-\eqref{c3}. Notice that the dual amplitudes depend on the point $(s_0,t_0,u_0)$ at which we evaluate the usual amplitude. Each dual amplitude is associated with one primal variable, namely with one spectral density and has a double jump on the physical regions depicted in green in fig.\ref{fig2}. After minimization, the double jump determines the physical partial waves from \eqref{a90} and \eqref{c55}.  

\section{Numerical implementation}
\label{sec5}
  
For the primal problem we have to compute the primal functional and the partial waves to impose unitarity. For that purpose, we map the region $4\le s \le \infty$ to $\xi\in[0,\pi]$ using
\beq
s=\frac{4}{\cos^2\frac{\xi}{2}} \;.
\label{a126}
\eeq  
Through a redefinition of the constant $f_0$ and spectral density $\sigma(x)$, we can rewrite the amplitude as
\beq
\begin{aligned}
	F(s,t,u) =& f_0 + \int_{0}^{\pi}\!\!\! d\xi_1 \,  \sigma(\xi_1) \left[ \cK(\xi_1,s) +\cK(\xi_1,t)+\cK(\xi_1,u) \right] 
	+ \int_{0}^{\pi}\!\!\! d\xi_1d\xi_2 \, \rho(\xi_1,\xi_2)\label{a127}\\ &\times\left[\cK(\xi_1,s)\cK(\xi_2,t)+\cK(\xi_1,s)\cK(\xi_2,u)+\cK(\xi_1,t)\cK(\xi_2,u)\right] \;,
\end{aligned}
\eeq
with
\beq
\cK(\xi,s) =\frac{1}{\pi} \frac{\sin\xi}{\frac{8}{s}-1-\cos\xi}\;.
\label{a128}
\eeq
To obtain this result we used
\beq
\intx\frac{\sigma(x)}{x-s} = \intx \frac{\sigma(x)}{x} + \intxi\ \cK(\xi,s)\ \sigma\left(\frac{4}{\cos^2\frac{\xi}{2}}\right)
\eeq
and renamed $\sigma\left(4\cos^{-2}\frac{\xi}{2}\right)\rightarrow \sigma(\xi)$. 
The first term is a constant that can be absorbed into $f_0$. Doing the same for $\rho(x,y)$ we get extra terms that can be absorbed into $\sigma(x)$ and $f_0$. 
To compute the partial waves it is useful to define
\begin{subequations}
	\beqa
	\Phi_{\ell}(s;\xi_1) &=&  \intP \cK(\xi,t) \label{a129}\\
	\tilde{\Phi}_{\ell}(s;\xi_{j_1},\xi_{j_2}) &=&  \intP \cK(\xi,t) \cK(\xi,u)  \label{a130}
	\eeqa
\end{subequations}
where, as usual, $t$ and $u$ are functions of $\mu$, $s$ as in \eqref{a4} and \eqref{a5}. 
Now that all functions are defined in a finite interval $[0,\pi]$ we can proceed to discretize the problem by evaluating the functions on a one-dimensional grid or by choosing a basis of functions. Choosing a basis leads to the method employed in \cite{Paulos:2017fhb} and we will discuss it later. First we consider the case of a discrete set of interpolation points since it gives a new alternative method that has some practical advantages.  

\subsection{Interpolation points}\label{inter}

 The most straight-forward approach is to discretize $\xi$ by choosing equally spaced points:
\beq
 \xi_j=j \Delta_\xi, \ \ \ \Delta_\xi = \frac{\pi}{M},\ \ 0\le j \le M , \;\  s_j=\frac{4}{\cos^2\frac{\xi_j}{2}}
 \label{a131}
\eeq
 The primal functional becomes
\beq
 \cF_P = a f_0 + a_{j_1} \sigma_{j_1} + a_{j_1j_2} \rho_{j_1 j_2}
\label{a132}
\eeq
with
\begin{subequations}
	\beqa
	a &=& 1    \label{a133}\\
	a_{j_1} &=& \Delta_\xi\, \big(\cK(\xi_{j_1},s_0)+\cK(\xi_{j_1},t_0)+\cK(\xi_{j_1},u_0) \big)   \label{a134}\\
	a_{j_1j_2} &=& \frac{\Delta^2_\xi}{2}\, \big(\cK(\xi_{j_1},s_0)\cK(\xi_{j_2},t_0)+\cK(\xi_{j_1},s_0)\cK(\xi_{j_2},u_0)\label{a135}\\
	&&+\cK(\xi_{j_1},t_0)\cK(\xi_{j_2},u_0) \big) +(j_1\leftrightarrow j_2)   \nonumber
	\eeqa
\end{subequations}
and the partial waves are
\beq
 h_{\ell j} = h_{\ell j} f_0 + h_{\ell j;j_1} \sigma_{j_1} + h_{\ell j;j_1j_2} \rho_{j_1 j_2} 
 \label{a136}
\eeq
 with 
 \begin{subequations}
 	\beqa
 	h_{\ell j} &=& \frac{\pi}{2} \sin\frac{\xi_j}{2} \delta_{\ell 0}    \label{a137}  \\
 	h_{\ell j;j_1}  &=& \frac{\pi}{2} \sin\frac{\xi_j}{2} (\delta_{\ell 0} \cK(\xi_{j_1},s^+_j) + \hat{\Phi}_{\ell}(s_j;\xi_{j_1}) )     \label{a138} \\
 	h_{\ell j;j_1j_2} &=& \frac{\pi}{2} \sin\frac{\xi_j}{2} \left[\cK(\xi_{j_1},s^+_j) \Delta_\xi \hat{\Phi}_{\ell}(s_j;\xi_{j_2}) +\half \Delta_\xi^2 \tilde{\Phi}_{\ell j; j_1 j_2}\right]\label{a139}\\
 	&& +(j_1\leftrightarrow j_2) \nonumber  
 	\eeqa
 \end{subequations}
 where $\cK(\xi_{j_1},s^+_j)$ should be understood as 
\beq
 \cK(\xi_{j_1},s^+_j) = \hat{K}_{jj_1} + \frac{i}{2\pi} \delta_{jj_1}
\eeq
 The matrix $\hat{K}_{j_1j_2}$ implements the principal part integral 
\beq
 f(\xi) = \frac{1}{\pi} \fint_0^\pi d\xi_1 \frac{\sin\xi_1}{\cos \xi-\cos\xi_1} g(\xi_1)
\eeq 
To implement this integral we extend $g(\xi_1)$ to the range $-\pi\le \xi_1\le \pi$ by assuming that it is anti-symmetric $g(-\xi_1)=- g(\xi_1)$ so that the integrand is symmetric. For the spectral densities $\sigma(\xi_1)$, $\rho(\xi_1,\xi_2)$ this agrees with them being the imaginary part of the amplitude that changes sign across the cut. Now we compute
\beq
 f(\xi_j) = \sum_{j_1=-M}^M K_{jj_1} g(\xi_{j_1})
\eeq 
where
\beq
 K_{j_1,j_2} = -\frac{1}{2M} (1-(-)^{j_1-j_2}) \cot\left(\frac{\pi}{2M}(j_1-j_2)\right)
\eeq
namely the discretized kernel that relates the real and imaginary part of an analytic function at the boundary of the unit disk. The dual problem can be constructed from this by discretizing the dual partial waves, namely using variables
\beq
 k_{\ell j} = k_\ell(\xi_j)  
 \label{a142}
\eeq
and computing the dual functional as
\beq
 \cF_D = \Delta_\xi \sum_{\ell,j} \left[ |k_{\ell j}| - \Re k_{\ell j} \right] + \Mreg ||\Theta||_* 
\label{a143}
\eeq
 For each primal variable there is a $\Theta$ component given by \eqref{a56}:
\begin{subequations}
	\beqa
	\Theta &=& a + \Delta_\xi \sum_{\ell j} \Im \left[\bar{k}_{\ell j} h_{\ell j} \right]  \label{a144}\\
	\Theta_{j_1} &=& a_{j_1} + \Delta_\xi \sum_{\ell j} \Im \left[\bar{k}_{\ell j} h_{\ell j;j_1} \right]   \label{a145}\\
	\Theta_{j_1j_2} &=& a_{j_1j_2} + \Delta_\xi \sum_{\ell j} \Im \left[\bar{k}_{\ell j} h_{\ell j;j_1j_2} \right]  \label{a146}
	\eeqa
\end{subequations}
where the coefficients are those in \eqref{a133}-\eqref{a139}. 
Now we require
\beq
 \Theta =0, \ \ \Theta_{j_1}=0
 \label{a147}
\eeq
which can be thought as taking $\Theta$ and $\Theta_{j_1}$ to have infinite norm unless they vanish. For the double dispersion we define 
\beq
 ||\Theta||_* = \sum_{\ell \j} \Im \left[h_{\ell j;j_1j_2}\right]  \Theta_{j_1j_2}
\label{a148}
\eeq
 In this way we have the dual of the numerical primal and therefore the minimum is equal to the maximum of the primal. It is a very useful way to evaluate the dual since it requires fewer partial waves. To obtain the actual dual that is above the maximum we have to restrict the space of dual partial waves. The way to accomplish this is to define a set of coefficients and expand the dual partial waves as
\beq
 k_\ell(\xi) = \sum_n k_{\ell n} e^{i n \xi}
\label{a149}
\eeq  
The coefficients $k_{ln}$ are taken real. Thus $k_\ell$ has a symmetric real part and antisymmetric imaginary part under $\xi \rightarrow -\xi$ as corresponds to the dual of a real analytic function as in \eqref{a55}. In our approach we evaluate the integrals numerically being careful to keep enough interpolation points to correctly describe the functions. As a practical rule we have that, if we have $N_{max}$ coefficients then we should have at the very least $M=4N_{max}$ interpolation points. So, the final form of the dual problem is 
\beq
\cF_D = \Delta_\xi \sum_j |\sum_n k_{\ell n} e^{i n \xi_j}| - \Delta_\xi \Re\left[ \sum_n k_{\ell n}  \sum_j e^{i n \xi_j}\right] + \Mreg ||\Theta||_* 
\label{a150}
\eeq
where
\beq
||\Theta||_* = \sum_{\ell jj_1j_2} \left| \Im h_{\ell j;j_1j_2}\right| \left|  a_{j_1j_2} + \sum_{\ell n} \Im[ k_{\ell' n} \theta_{\ell' n;j_1 j_2}] \right| 
\label{a151}
\eeq
and we further impose the constraints 
\begin{subequations}
	\beqa
	0 &=& a + \sum_{\ell n} \Im[ k_{\ell n} \theta_{\ell n}]  \label{a152} \\
	0 &=& a_{j_1} +  \sum_{\ell n} \Im[ k_{\ell n} \theta_{\ell n;j_1}] \label{a153}
	\eeqa
\end{subequations}
The coefficients $\theta^{(a)}$ are given by 
\begin{subequations}
	\beqa
	\theta_{\ell n} &=&  \sum_{j} e^{i n j} h_{\ell j} \label{a154} \\
	\theta_{\ell n;j_1} &=&  \sum_{j} e^{i n j} h_{\ell j;j_1}  \label{a155} \\
	\theta_{\ell n;j_1j_2} &=&  \sum_{j} e^{i n j} h_{\ell j;j_1j_2} \label{a156}
	\eeqa
\end{subequations}
which can be efficiently evaluated using the Fast Fourier Transform.

\subsection{Discrete basis of functions}\label{basis}

It is useful both conceptually and also for numerical purposes to consider the same problem in a discrete basis of functions by taking
\begin{subequations}
\beqa
\rho(x,y) &=& \sum_{n_1,n_2} \rho_{n_1n_2} \phi_{n_1}(x) \phi_{n_2}(y) \label{a157}\\
\sigma(x) &=& \sum_n \sigma_n \phi_n(x)  \label{a158}
\eeqa
\end{subequations}
The simplest functions to use as a basis are ($n\ge 1$)
\beq
\phi_n (x) =\sin(n\varphi) , \ \ \ \ x= \frac{4}{\cos^2\frac{\varphi}{2}} 
\label{a159}
\eeq
satisfying $\phi_n(x\rightarrow 4)\sim \sqrt{x-4}$ and $\phi_n(x)\simeq x^{-1/2}$ for $x\rightarrow\infty$. Using Chebyshev polynomials we can also write $\phi_n(x) = 4\frac{\sqrt{x-4}}{x} U_{n-1}(1-\frac{8}{x})$. The analytic functions $\Phi_n(s)$ are:\footnote{The constant $(-1)^n$ can be omitted.}
\beq
\Phi_{n}(s) = \frac{1}{\pi} \int_4^\infty\!\!\!\!\!\! dx\ \frac{\phi_{n}(x)}{x-s}= z^n-(-1)^n 
\label{a160}
\eeq
with
\beq
z = \frac{2-\sqrt{4-s}}{2+\sqrt{4-s}}, \ \ \ \ z(x+i\epsilon)=e^{i\varphi}, 
\label{a161}
\eeq
To this set it is convenient to add an extra function $\phi_0(x)=\frac{1}{\sqrt{x-4}}$ that allows the inclusion of a divergence at $x=4$ (or $y=4$). Here we only do that as a check since the purpose is to obtain the existence of a pole from the optimization.  We then have
\beqa
F(s,t,u) &=& f_0 + \sum_n \sigma_n\, \left[\Phi_{n}(s) +\Phi_{n}(t)+\Phi_{n}(u)\right] \label{a162} \\ 
&& + \sum_{nm} \rho_{nm}\ \left[\Phi_{n}(s) \Phi_{m}(t)+\Phi_{n}(s) \Phi_{m}(u)+\Phi_{n}(t) \Phi_{m}(u)\right] , \nonumber \label{a163} 
\eeqa
Let us define 
\begin{subequations}
	\beqa
	\hat{\Phi}_{n,\ell}(s) &=& \int_{-1}^{+1} d\mu P_\ell(\mu) \Phi_{n}(t)  \label{a164}\\
	\tilde{\Phi}_{nm,\ell}(s) &=& \int_{-1}^{+1} d\mu  P_\ell(\mu) \Phi_{n}(t)\Phi_{m}(u)  \label{a165}
	\eeqa
\end{subequations}
where as always, in the integrand, $t$ and $u$ are functions of $\mu,s$ as in \eqref{a4}, \eqref{a5}. Now we can write the partial waves ($\ell$ even) as
\beq
h_\ell(s) =  A_\ell(s) f_0 + \sum_{n} A_{\ell,n}(s)\ \sigma_{n} + \sum_{nm} A_{\ell,nm}(s)\ \rho_{nm}, \label{a166}
\eeq
with
\begin{subequations}
	\beqa
	A_\ell(s) &=& \frac{\pi}{2}\sqrt{\frac{s-4}{s}} \delta_{\ell0} \\
	A_{\ell,n}(s) &=&  \frac{\pi}{2}\sqrt{\frac{s-4}{s}}  \left( \Phi_{n}(s^+) \delta_{\ell0} + \hat{\Phi}_{n,\ell}(s) \right)   \label{a167} \\
	A_{\ell,nm}(s) &=& \frac{\pi}{4}\sqrt{\frac{s-4}{s}}  \left( \Phi_{n}(s^+) \hat{\Phi}_{m,\ell}(s)+ \half  \tilde{\Phi}_{nm,\ell}(s) \right)\\
	 &&+(n\leftrightarrow m)\nonumber
	\label{Asp}
	\eeqa
\end{subequations}
The functional is 
\beq
\cF_P = a_0 f_0 + \sum_{n} a_n \sigma_n + \sum_{nm} a_{nm} \rho_{nm}
\label{a168} 
\eeq
For example if we take $\cF_P=F(s_0,t_0,u_0)$ then 
\beqa
a_0 &=& 1 \ , \\
a_n &=& \Phi_{n}(s_0) + \Phi_{n}(t_0)+\Phi_{n}(u_0) \ , \label{a169}\\
a_{nm} &=& \half\left[ \Phi_{n}(s_0) \Phi_{m}(t_0)+\Phi_{n}(s_0) \Phi_{m}(u_0)+\Phi_{n}(t_0) \Phi_{m}(u_0)\right]\\
 &&+ (n\leftrightarrow m) \ , \non
\label{fsp}
\eeqa
The dual problem is
\beq
\min\left\{\cF_D =\intlevens (|k_{\ell}(s)| - \Re k_{\ell}(s) ) + \frac{M_{\mbox{reg}}}{\pi^2} ||\Theta||_*  \right\} \ ,
\label{a170}
\eeq
over the space of complex functions of a real variable $k_\ell(s)$ ($s\in\mathbb{R}_{>4}$) with the constraints
\begin{subequations}
	\beqa
	1 &=& \int_4^\infty\!\!\!\! ds\, f_0(s)\, \Im[\bar{k}_0(s)]  \label{a173} \\
	a_n &=& \int_4^\infty\!\!\!\! ds\, f_0(s) \left\{\Im[\bar{k}_0(s) \Phi_n(s^+)]+ \sumeven \Im[\bar{k}_\ell(s)]\, \hat{\Phi}_{n,\ell}(s) \right\} , \label{a174}
	\eeqa
\end{subequations}
which come from setting to zero the coefficients $\Theta$ associated with $f_0$ and $\sigma_n$. For the double dispersion relation we have 
\beqa
\Theta_{nm} &=& f_{nm}+\pi^3\intlevens \sqrt{\frac{s-4}{s}} \ \Im \left[ k_\ell(s) A_{\ell,nm}(s) \right] \ , \label{a171} \\
||\Theta||_*^2 &=& \sum_{nm} \left(\Theta_{nm}\right)^2 \ ,  \label{a172}
\eeqa
where we used the usual square norm but others lead to the same results. The constraints \eqref{a174} are a small set of contraints so the space of functions $k_\ell(s)$ is largely unconstrained. The full set of dual constraints should include $\Theta_{nm}=0$ that, as mentioned cannot be imposed since it has no solutions. However, if $M_{\mbox{reg}}\gg 1$, minimizing the functional generically results in functions $k_\ell(s)$ such that $\Theta_{nm} \simeq 0$.

In the $O(N)$ isospin case, we proceed in the same way. The $(t\leftrightarrow u)$ symmetric amplitude $A(s,t,u)$ in \eqref{a8} is written as
\beqa
 A(s,t,u) &=& f_0 + \sum_n \left[\sigma^{(1)}_n \Phi_n(s) + \sigma^{(2)}_n (\Phi_n(t)+\Phi_n(u))\right] \label{d1} \\
          && +\sum_{nm} \left[\rho^{(1)}_{nm} \Phi_n(s) (\Phi_m(t)+\Phi_m(u)) + \rho^{(2)}_{nm} \Phi_n(t)\Phi_m(u) \right] \non
\eeqa
using the same basis functions as in \eqref{a162}. The primal variables are the $\sigma^{(1,2)}_n$, $\rho^{(1,2)}_{nm}$ with the only condition that $\rho^{(2)}_{nm}=\rho^{(2)}_{mn}$ is symmetric. 
The partial waves are 
\beqa
 h^I_\ell(s) &=& \frac{\pi}{4} \sqrt{\frac{s-4}{s}} \left\{f_0 A^I_\ell + \sum_n \left[ A^I_{\ell n}(s)\sigma^{(1)}_n + B^I_{\ell n}(s) \sigma^{(2)}_n \right] \right.   \label{a175} \\
        && \left. + \sum_{nm} \left[ A^I_{\ell,nm}(s)\ \rho^{(1)}_{nm} + B^I_{\ell,nm}(s)\ \rho^{(2)}_{nm}\right] \right\} \ ,  \nonumber \label{a176}
\eeqa
where (we omit the argument $s$ in all functions):
\beq
\begin{aligned}
A^0_\ell &= 2(N+2)\delta_{\ell 0},  & A^1_\ell &= 0, & A^2_\ell &= 4\delta_{\ell 0}, \\
A^0_{\ell,n} &= 2N\delta_{\ell 0}\Phi_n+ 2\hat{\Phi}_{n\ell}, & A^1_{\ell,n} &= 2 \hat{\Phi}_{n\ell}, & A^2_{\ell,n} &= 2 \hat{\Phi}_{n\ell}, \\
B^0_{\ell,n} &= 4\delta_{\ell 0} \Phi_n + 2(N+1) \hat{\Phi}_{n\ell},  & B^1_{\ell,n} &=  -2\hat{\Phi}_{n\ell},  & B^2_{\ell,n} &= 4 \delta_{\ell 0} \Phi_n + 2 \hat{\Phi}_{n\ell}, 
\end{aligned}
\eeq
and
\begin{subequations}
\beqa
 A^0_{\ell,nm} &=& 2N\Phi_n\hat{\Phi}_{m\ell}+ 2\Phi_m\hat{\Phi}_{n\ell}+2\tilde{\Phi}_{nm,\ell}   \ ,\label{a186} \\
 A^1_{\ell,nm} &=& 2 \Phi_m\hat{\Phi}_{n\ell}+2\tilde{\Phi}_{nm,\ell}   \label{a187} \ ,\\ 
 A^2_{\ell,nm} &=& 2 \Phi_m\hat{\Phi}_{n\ell}+2\tilde{\Phi}_{nm,\ell}   \label{a188} \ ,\\
 B^0_{\ell,nm} &=&  \Phi_n\hat{\Phi}_{m\ell}+\Phi_m\hat{\Phi}_{n\ell}+ N\tilde{\Phi}_{nm,\ell} \ ,\label{a189} \\
 B^1_{\ell,nm} &=&  -\Phi_n\hat{\Phi}_{m\ell}-\Phi_m\hat{\Phi}_{n\ell} \label{a190} \ ,\\
 B^2_{\ell,nm} &=&  \Phi_n\hat{\Phi}_{m\ell}+\Phi_m\hat{\Phi}_{n\ell}  \label{a191} \ ,
 \eeqa
\end{subequations}
where for $I=0,2$ we only have $\ell$ even and for $I=1$ only $\ell $ odd. Finally we choose a functional
\beq
 \cF_P =a_0 f_0 + \sum_n f_n \sigma^{(1)}_n + g_n \sigma^{(2)}_n + \sum_{nm} f_{nm} \rho^{(1)}_{nm}+g_{nm} \rho^{(2)}_{nm} \ .
\label{a192} 
\eeq
The dual problem is now
\begin{equation}
\begin{aligned}
\mmin{k_\ell^I(s)}\Big\{F_D &=\sum_{(I,\ell)} \int_4^\infty\!\!\!\!\!\! ds (|k^I_{\ell}(s)| + \Re k^I_{\ell}(s) ) + M_{\mbox{reg}} \left(||\Theta||_*+||\Xi||_*\right)  \Big\} \ ,\\
\Theta_{nm} &= f_{nm}+\frac{\pi}{4} \intlsI  \sqrt{\frac{s-4}{s}} \Im\left[ k^I_\ell(s) A_{\ell,nm}(s) \right] \ ,\\
\Xi_{nm}    &= g_{nm}+\frac{\pi}{4} \intlsI \sqrt{\frac{s-4}{s}} \Im\left[ k^I_\ell(s) B_{\ell,nm}(s) \right] \ ,
\label{a193}
\end{aligned}
\end{equation}
subject to the constraints associated with the constant $f_0$ and the single dispersion relation
\begin{subequations}
	\beqa
	\Theta   &=& a_0 + \frac{\pi}{4}\intlsI \sqrt{\frac{s-4}{s}} \Im\left[ \bar{k}_\ell^I(s) A^I_\ell \right] =0 \ ,\\
	\Theta_n &=& f_n + \frac{\pi}{4}\intlsI \sqrt{\frac{s-4}{s}} \Im\left[ \bar{k}_\ell^I(s) A^I_{\ell n}(s) \right] =0  \ , \\
	\Xi_n    &=& g_n + \frac{\pi}{4}\intlsI \sqrt{\frac{s-4}{s}} \Im\left[ \bar{k}_\ell^I(s) B^I_{\ell n}(s)\right] = 0  \ .
	\eeqa
\end{subequations}
Not imposing these constraints would be the dual to the problem where $f_0=0$ and $\sigma^{(1,2)}_n=0$, namely only with double dispersion relation.
It is useful to notice that, in constructing the dual problem we can directly use the formulas \eqref{a63} since they do not depend on how we choose the variables $\alpha_n$ to define the primal problem.  
Notice that once again in the dual problem the functions $k^I_{\ell}(s)$ are independent. Crossing only appears in the dual problem through the coefficients $A_{\ell,nm}(s)$, $B_{\ell,nm}(s)$.
Finally, the dual problem thus defined is the dual of the numerical primal problem and then its minimum should agree with the maximum of the primal. To obtain the actual convex dual we have to restrict the space of variables $k_\ell(s)$ by introducing once again a set of real coefficients to parameterize the dual variables:
\beq
 k_\ell(\xi) = \sum_p k_{\ell p} e^{i p \xi}\ .
\label{c149}
\eeq  
 The number of coefficients $k_{\ell p}$ should be much smaller than the number of coefficients in representing the amplitude \eqref{a159}. The reason is that now, those coefficients are taken as a way to evaluate the integrals in the dual problem and cannot do that accurately if the functions $k_\ell(s)$ fluctuate more rapidly than the functions in the basis \eqref{a159}.

\section{A few numerical results}
\label{sec6}

The main idea for the numerical approach is that, in the primal we parameterize the amplitude using the interpolation points as described in \ref{inter}, or as linear combinations of a set of functions as in \ref{basis}. As we increase the size of this set, the numerical maximum increases approaching the true maximum. In the dual, we write the dual partial waves as linear combinations of a set of functions and compute the minimum. As we increase the size of the set, the minimum decreases approaching the maximum of the primal problem from above. To check the method, we choose a known but challenging problem: maximizing the value of the amplitude for single pions at a point in the Mandelstam triangle, for example at the symmetric point $s_0=t_0=u_0=4/3$. This is numerically challenging because the maximum is attained by a function that has a pole at threshold ($s=4$) as studied in \cite{Paulos:2017fhb}. There, it was noticed that the convergence to the maximum in terms of the size of the basis is slow unless one explicitly includes a pole at threshold. For that reason the dual problem we study here is important since it provides an upper bound that is above the maximum. Note that this is all done {\em without assuming the existence of a pole at threshold}. In fact one can use the same no-pole coefficients for the primal and dual. Of course, although the coefficients are the same, they enter differently in the primal and the dual leading to lower and upper bounds. For the actual optimization we use standard software \cite{cvx,gb08}.
\begin{figure}
	\centering
	\includegraphics[width=\textwidth]{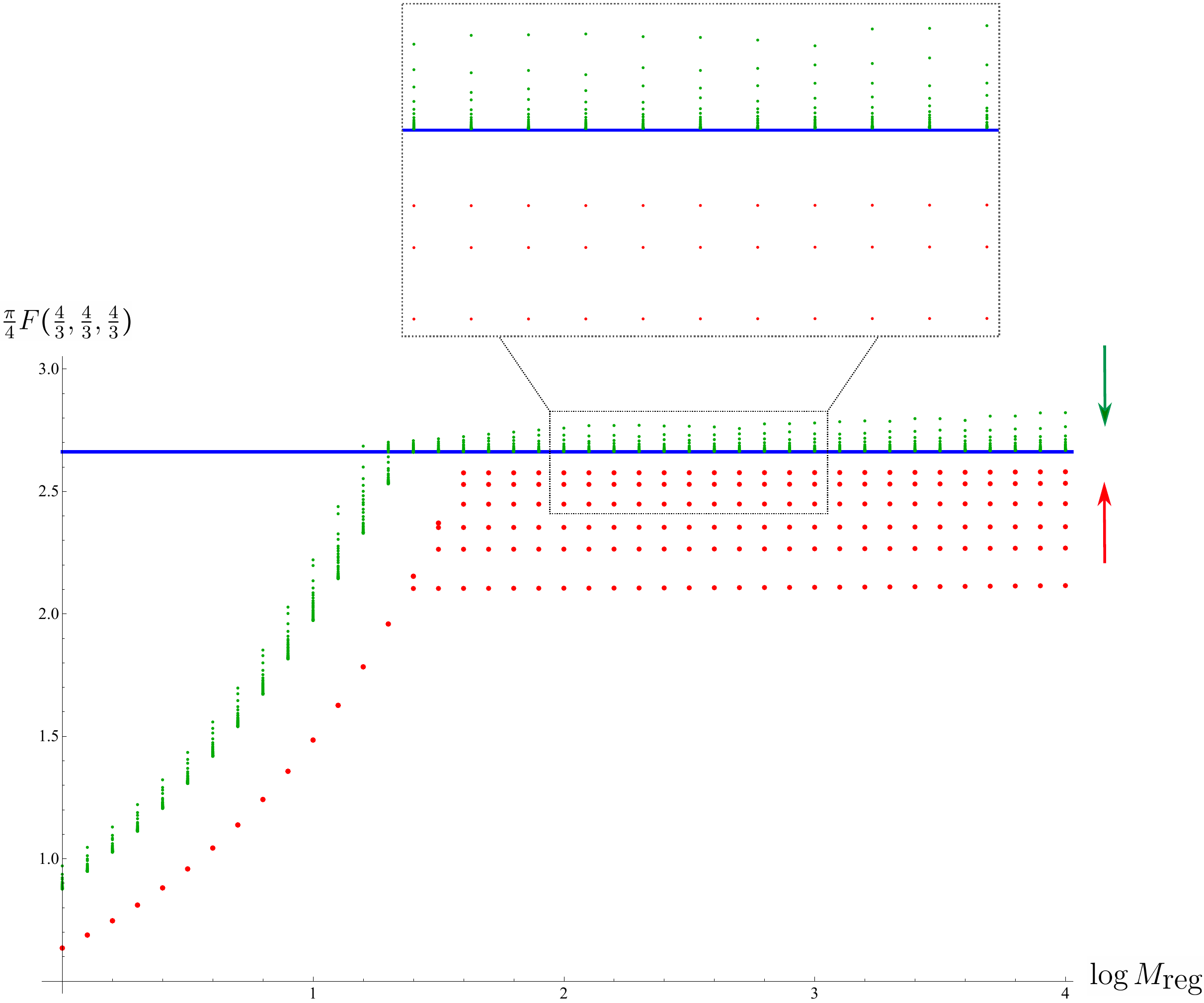}
	\caption{The maximum of the amplitude at the symmetric point $s_0=t_0=u_0=\frac{4}{3}$ as the regulator $\Mreg$ is increased for the primal (red) and dual (green) problems. The horizontal axis gives the regulator in a logarithmic scale and up to a horizontal shift to plot the curves together. As explained in the main text, the primal improves by increasing the number of interpolation points, as indicated with the red upward arrow, and the dual improves by increasing the number of coefficients, as indicated with the green downward arrow. We do not asume the existence of a pole at threshold which would give the horizontal blue line. The values are multiplied by a factor $\pi/4$ to match the normalization in the literature, \eg\ \cite{Paulos:2017fhb}.}
	\label{fig1}
\end{figure}
\begin{figure}
	\centering
	\includegraphics[width=\textwidth]{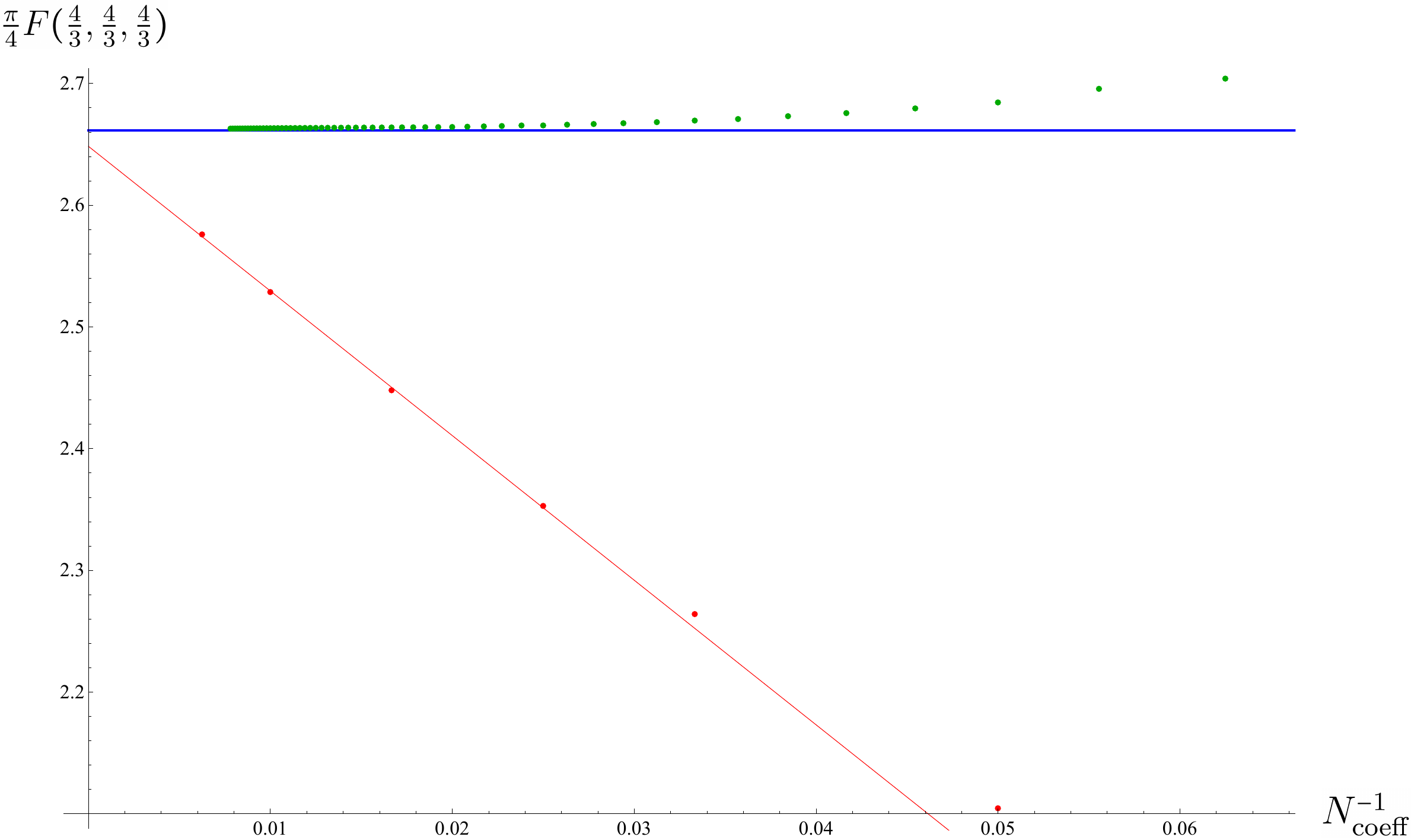}
	\caption{The maximum of the amplitude at the symmetric point $s_0=t_0=u_0=\frac{4}{3}$ as the (inverse of the) number of interpolation points/coefficients $N^{-1}_{\text{coeff}}$ changes for the primal (red) and the dual (green). The vertical axis correspond to infinite number of interpolation points for the primal or coefficients for the dual. We do not assume the existence of a pole at threshold which would give the horizontal line. The values are multiplied by a factor $\pi/4$ to match the normalization in the literature, \eg\ see \cite{Paulos:2017fhb}.}
	\label{fig2}
\end{figure}

\begin{figure}
	\begin{centering}
		\begin{subfigure}[h]{\textwidth}
			\centering
			\includegraphics[width=0.85\textwidth]{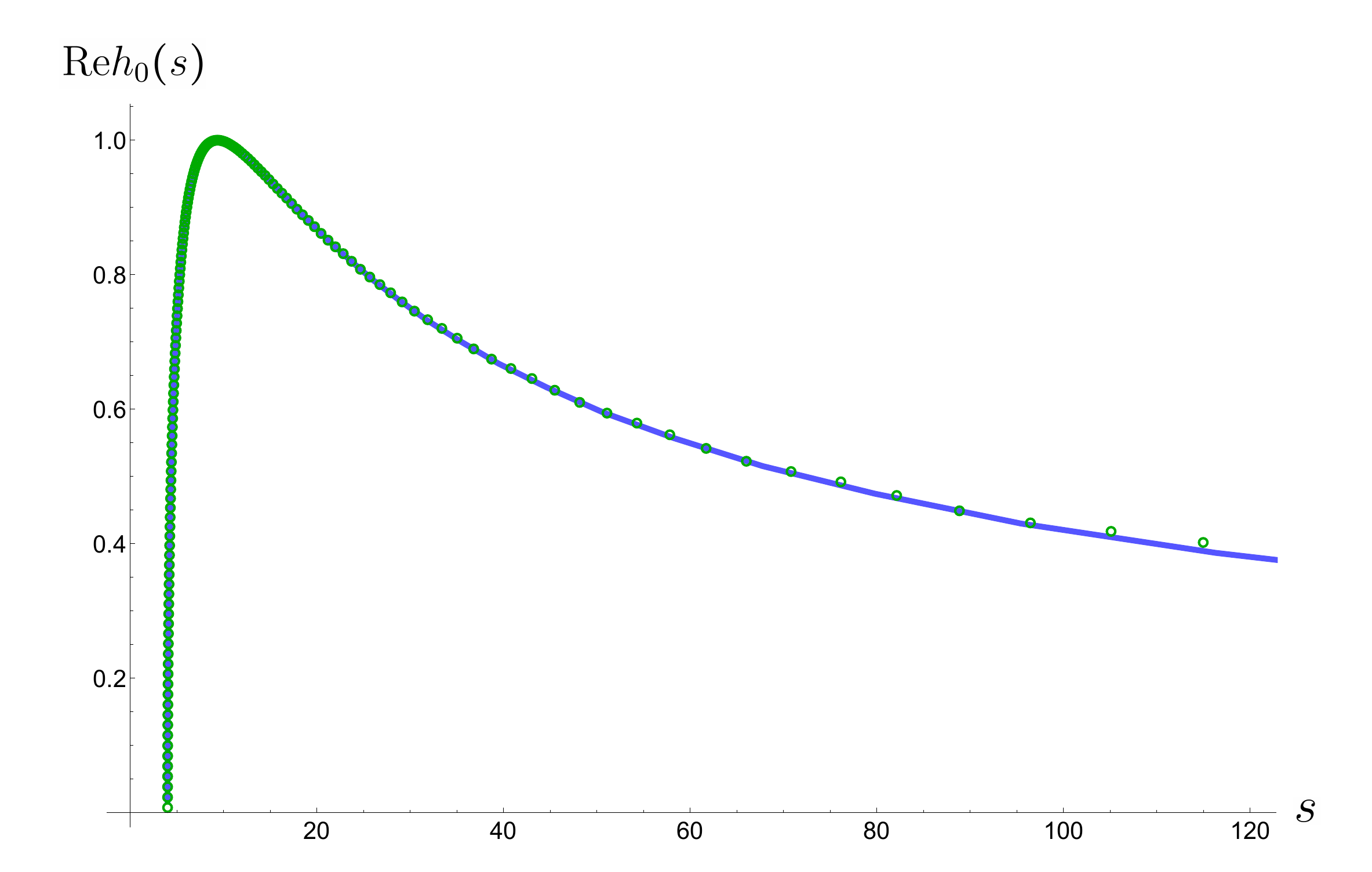}
			\caption{}
		\end{subfigure}
		\begin{subfigure}[h]{\textwidth}
			\centering
			\includegraphics[width=0.85\textwidth]{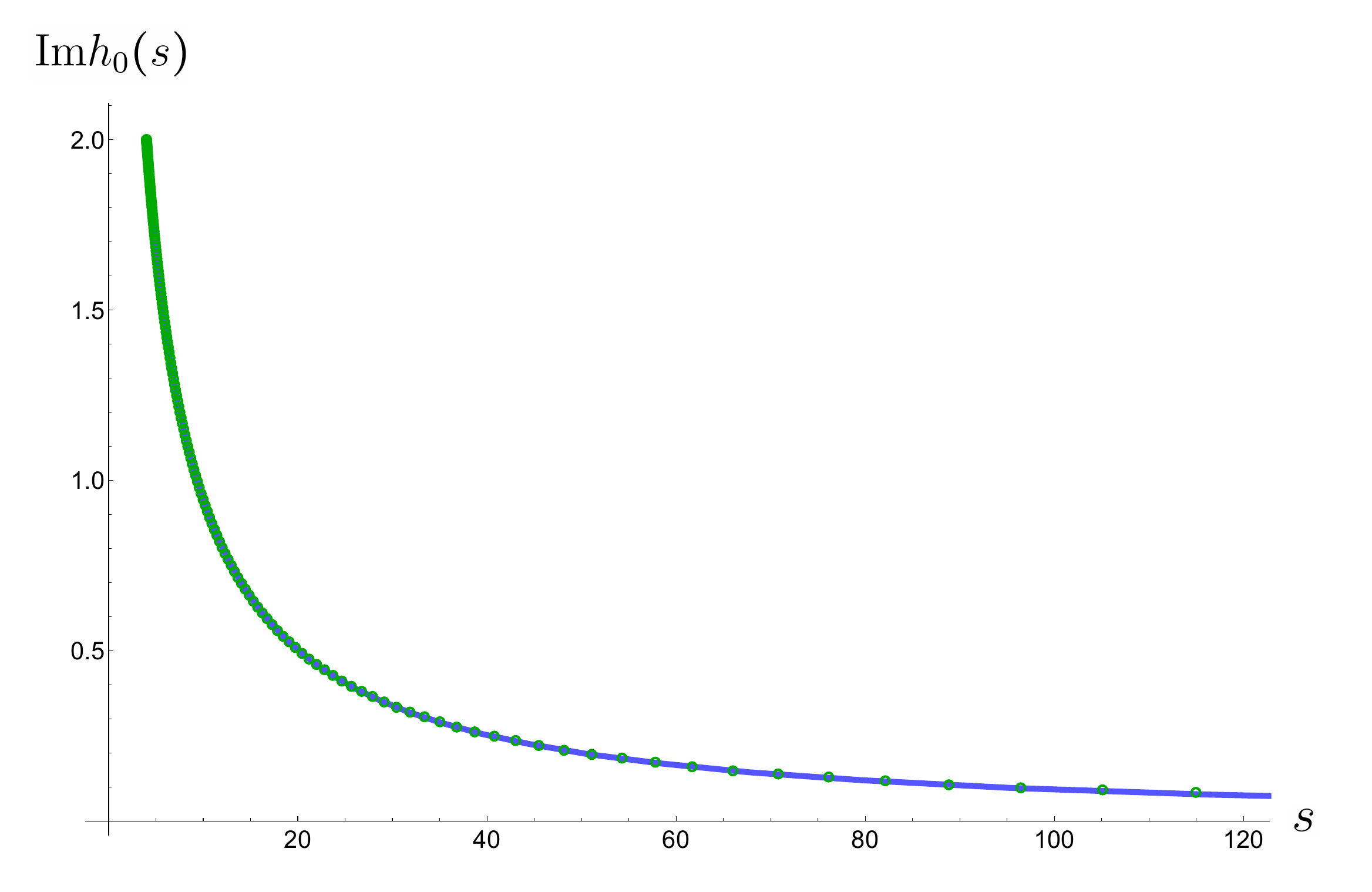}
			\caption{}
		\end{subfigure}
	\end{centering}
	\caption{Real and imaginary parts of the S-wave $h_0(s)$ for the amplitude from maximizing at the symmetry point. The blue curve is the primal result by imposing a threshold pole, and the green circles are the dual result using the expression \eqref{c55} {\it without} assuming the threshold pole. The two agree perfectly.}
	\label{ReImh0}
\end{figure}

\begin{figure}
	\centering
	\includegraphics[width=\textwidth]{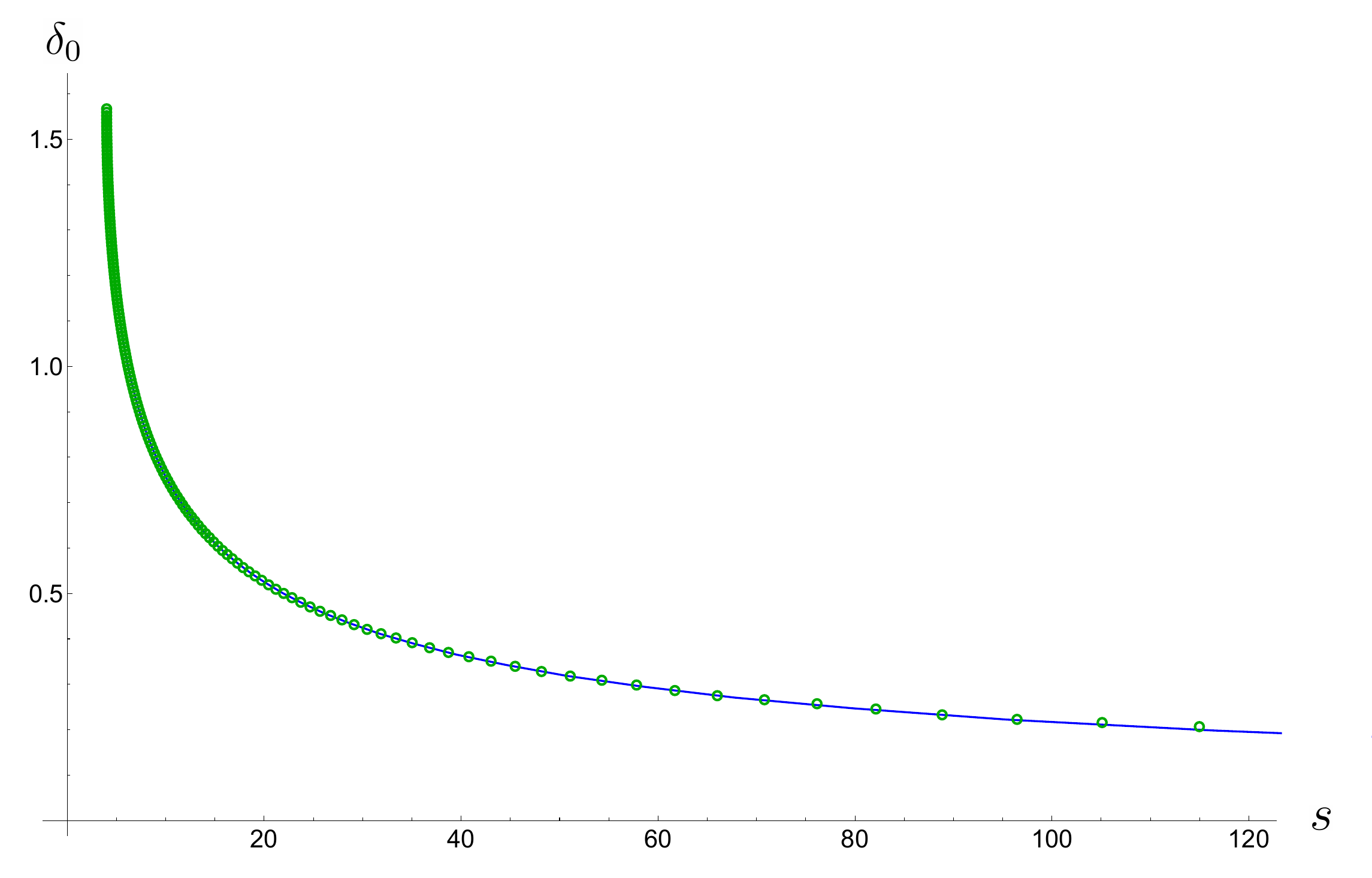}
	\caption{The phase shift $\delta_0(s)$ of the S-wave, defined in eq. \eqref{Sldef} for the amplitude from maximizing at the symmetry point. The blue curve is the primal result by imposing a threshold pole, and the green circles are the dual result using the expression \eqref{c55} {\it without} assuming the threshold pole. The two agree perfectly.}
	\label{deltaS}
\end{figure}

We considered first the amplitude at the symmetric point $\cF=F(\frac{4}{3},\frac{4}{3},\frac{4}{3})$. In fig.\ref{fig1} we see how, increasing the regulator $\Mreg$ the maximum increases until it reaches a plateau that cover several orders of magnitude of the regulator (notice the logarithmic scale in the horizontal axis). That plateau is taken as the maximum. For larger values of the regulator the maximum can increase further or fluctuate randomly due to numerical errors, that part should be ignored. The dual problem reaches the plateau always above the actual maximum and the primal always below. As shown in fig. \ref{fig1}, in the primal problem we used $M=20, 30, 40, 60, 100, 160$ interpolation points for the single and double dispersion relations where unitarity constraints are imposed.
For the dual we used $512$ interpolation points and $16 \le N_{\mathrm{coeff}} \le 128$ and only dual partial waves up to $\ell_{\mathrm{max}}=4$. It is interesting to extrapolate the result to a very large number of interpolation points or coefficients. This is done in fig.\ref{fig2} where one can see that for an infinite number of coefficients (vertical axis) the maximum converges to a number very close to the value obtained assuming the existence of a pole. The dual seems to be doing better without needing extrapolation. 

At the plateau of figure \ref{fig3}, we can compute the physical partial waves $h_\ell(s)$ from the dual partial waves $k_{\ell}(s)$ using eq. \eqref{a55}. As mentioned before, the dual formulation does not assume the threshold pole as the maximum is attained, and can therefore detect such a pole as a result of the optimization. The corresponding phase shift $\delta_{\ell}(s)$ can then be computed using the definition \eqref{Sldef}. In figs. \ref{ReImh0} and \ref{deltaS}, we show the S-wave $h_0(s)$ and its phase shift $\delta_0(s)$ obtained from the dual partial waves and compare it with the primal result when a pole is included in the setup, and find perfect agreement.

To show the power of the method further,  we consider $\cF=\max F(s_0,s_0,4-2s_0)$, with $0<s_0<4$ whose result have been displayed in fig.\ref{fig3} in section \ref{intro}.  The primal data points were obtained by using basis functions as described in section \ref{basis}. The lower curve is done with $8\times 8$ coefficients for the double dispersion, $100$ coefficients for the single dispersion and checking unitarity up to $\ell_\mathrm{max}=10$ at $100$ values of $s$. The upper red curve is $10\times 10$ coefficients for double dispersion and $300$ for single, checking unitarity at $300$ points for $\ell_\mathrm{max}=20$. To check we also run $100$ interpolation points (for single and double dispersion relation) which agreed with the lower curve. This is because the pole at threshold is in the s-wave and can be reproduced from the single dispersion relation with enough coefficients. In general running $100$ interpolation points should be better. The dual (green points) was run using 512 interpolation points and using partial waves up to $\ell_{\mathrm{max}}=12$. Finally, 
the blue line is the result of including a pole\footnote{In fact, from the numerical results that include a pole, it appears that the amplitude is the same along the curve, that is, the maximum is attained at all these points by the same function. However we were not able to derive this fact analytically so we present it as an interesting idea for future work suggested by the numerics. We thank the referee for pointing out this possibility.}. We can see that the primal and the dual bracket that line. Even if we did not have the blue line, we are still assured that the program is converging from above and below. The black diamonds are the pioneering results of Lopez and Mennessier in \cite{LM} where upper bounds were found by a somewhat different method. When $s_0=t_0\rightarrow 0$ and $s_0=t_0\rightarrow 4$, numerically we observe a divergence $\cF_{\mathrm{max}}\sim \frac{1}{\sqrt{s_0}}$ and $\cF_{\mathrm{max}}\sim \frac{1}{\sqrt{4-s_0}}$ respectively and we also find a minimum at the symmetric point $s_0=t_0=u_0=\frac{4}{3}$ where $\frac{\pi}{4}F(\frac{4}{3},\frac{4}{3},\frac{4}{3})\simeq 2.6613$. This value at the symmetric point agrees with the result in \cite{Paulos:2017fhb}, we just observe that it is the minimum of the blue curve.

\section{Conclusions}\label{con}

 The S-matrix bootstrap provides a method to investigate the space of S-matrices allowed by the general constraints of analyticity, unitarity, crossing, and global symmetries. When restricted to a subsector such as $2\rightarrow 2$ scattering the space can be efficiently mapped by using conic maximization numerical tools. In particular the space is convex. At the boundary of the space sometimes distinguished points such as vertices can be found that define interesting S-matrices which correspond to physically important theories. Since the numerics requires discretizing the problem, an approximation is made. One can control the approximation by running a primal problem that reduces the number of variables on which the S-matrix depends. In that case we map an interior region of the actual allowed space that grows as we increase the number of variables. Alternatively, one can do an approximation in the dual convex problem, in which one restricts the space of dual variables, namely of Lagrange multipliers that enforce the constraints. In that case one obtains an exterior region that contains the allowed region of S-matrices. By running both problems we bracket the region between an interior and an exterior one \cite{Cordova:2019lot}. This is particularly useful in the case of a challenging numerical problem such as the scattering of scalar particles in $3+1$ dimensions. The problem is that the maximum is attained by an amplitude that has a pole at threshold \cite{Paulos:2017fhb}. One would like to reproduce those results without assuming the existence of a pole. In this paper we constructed explicitly the dual problem for scattering of scalars in $3+1$ dimensions, both  for a single scalar and the more general $O(N)$ model. Numerically, when applied to the single scalar case we obtain a good bracketing of the maximum without assuming the existence of a pole. Another interesting property is that the dual problem allows for the possibility of unitarity unsaturation, however, in practice, it is not clear when that would happen. In summary, this work provides a new valuable tool that can be efficiently used to map out the space of S-matrices in higher dimensional theories. 

\section{Acknowledgements}
 We are very grateful to Lucia Cordova, Harish Murali, Joao Penedones and Pedro Vieira for discussions, as well as comments and suggestions on the draft. We would also like to thank the referee from JHEP for many interesting questions and suggestions. In addition, M.K. is very grateful to the DOE that supported in part this work through grants DE-SC0007884, DE-SC0019202 and the QuantiSED Fermilab consortium, as well as to the Keck Foundation that also provided partial support for this work.

\appendix

\numberwithin{equation}{section}

\section{Useful formulas}
\label{appendix}

In \eqref{a27} and \eqref{a28}, using results from \cite{gradshteyn2007}, the integrals over $\mu$ were done analytically in terms of the Legendre functions $Q_\ell$  using
\beqa
\int_{-1}^{+1}\!\!\!\!\!\! d\mu\,  \frac{P_\ell(\mu)}{x-t} &=& \frac{4}{s-4} \Qx  \label{a194}\\
\int_{-1}^{+1}\!\!\!\!\!\! d\mu\, \frac{P_\ell(\mu)}{(x-t)(y-u)} &=& \frac{4}{s-4} \frac{\Qx+(-)^\ell \Qy}{ (s-4+x+y)}  \label{a195}
\eeqa
where, $t$ and $u$ are functions of $\mu$ and $s$ given by \eqref{a4} and \eqref{a5}. Another useful formula is \cite{gradshteyn2007}
\beq
 \sum_{\ell=0}^\infty (2\ell+1) P_\ell(x_1) Q_\ell(x_2) = \frac{1}{x_2-x_1}, \ \ |x_1+\sqrt{x_1^2-1}|< |x_2+\sqrt{x_2^2-1}|
 \label{a196}
\eeq
 It is interesting to notice that, using this last identity one can see that
\beq
 k_\ell(s) = -\frac{2}{\pi^2} \sqrt{\frac{s}{s-4}}\left(\frac{1}{s-t_0}+\frac{1}{s-u_0}\right)\, P_\ell\left(1+\frac{2s_0}{s-4}\right)
\label{a197}
\eeq
 solves the constraint $\Theta(x,y)=0$ in \eqref{a57}. However the argument of the Legendre polynomial is larger than one and therefore the functions $k_\ell(s)$ are real and negative and grow in absolute value with $\ell$. Therefore the functional $\cF_D$ in \eqref{a56} is divergent. Also the sum in \eqref{a91} is divergent and \eqref{a197} does not lead to a well defined dual amplitude in agreement with the observation after  \eqref{a116}.

\bibliographystyle{utphys}
\bibliography{references}

\providecommand{\href}[2]{#2}\begingroup\raggedright\begin{thebibliography}{10}

\bibitem{Eden:1966dnq}
R.~J. Eden, P.~V. Landshoff, D.~I. Olive, and J.~C. Polkinghorne, {\em {The
  analytic S-matrix}}.
\newblock Cambridge Univ. Press, Cambridge, 1966.

\bibitem{chew1966analytic}
G.~Chew, {\em The Analytic S Matrix: A Basis for Nuclear Democracy}.

\bibitem{Paulos:2016fap}
M.~F. Paulos, J.~Penedones, J.~Toledo, B.~C. van Rees, and P.~Vieira, ``{The
  S-matrix bootstrap. Part I: QFT in AdS},''
  \href{http://dx.doi.org/10.1007/JHEP11(2017)133}{{\em JHEP} {\bfseries 11}
  (2017) 133}, \href{http://arxiv.org/abs/1607.06109}{{\ttfamily
  arXiv:1607.06109 [hep-th]}}.

\bibitem{Paulos:2016but}
M.~F. Paulos, J.~Penedones, J.~Toledo, B.~C. van Rees, and P.~Vieira, ``{The
  S-matrix bootstrap II: two dimensional amplitudes},''
  \href{http://dx.doi.org/10.1007/JHEP11(2017)143}{{\em JHEP} {\bfseries 11}
  (2017) 143}, \href{http://arxiv.org/abs/1607.06110}{{\ttfamily
  arXiv:1607.06110 [hep-th]}}.

\bibitem{Paulos:2017fhb}
M.~F. Paulos, J.~Penedones, J.~Toledo, B.~C. van Rees, and P.~Vieira, ``{The
  S-matrix bootstrap. Part III: higher dimensional amplitudes},''
  \href{http://dx.doi.org/10.1007/JHEP12(2019)040}{{\em JHEP} {\bfseries 12}
  (2019) 040}, \href{http://arxiv.org/abs/1708.06765}{{\ttfamily
  arXiv:1708.06765 [hep-th]}}.

\bibitem{Homrich:2019cbt}
A.~Homrich, J.~a. Penedones, J.~Toledo, B.~C. van Rees, and P.~Vieira, ``{The
  S-matrix Bootstrap IV: Multiple Amplitudes},''
  \href{http://dx.doi.org/10.1007/JHEP11(2019)076}{{\em JHEP} {\bfseries 11}
  (2019) 076}, \href{http://arxiv.org/abs/1905.06905}{{\ttfamily
  arXiv:1905.06905 [hep-th]}}.

\bibitem{Polyakov:1975rr}
A.~M. Polyakov, ``{Interaction of Goldstone Particles in Two-Dimensions.
  Applications to Ferromagnets and Massive Yang-Mills Fields},''
  \href{http://dx.doi.org/10.1016/0370-2693(75)90161-6}{{\em Phys. Lett. B}
  {\bfseries 59} (1975) 79--81}.

\bibitem{Zamolodchikov:1977nu}
A.~B. Zamolodchikov and A.~B. Zamolodchikov, ``{Relativistic Factorized S
  Matrix in Two-Dimensions Having O(N) Isotopic Symmetry},''
  \href{http://dx.doi.org/10.1016/0550-3213(78)90239-0}{{\em JETP Lett.}
  {\bfseries 26} (1977) 457}.

\bibitem{Zamolodchikov:1978xm}
A.~B. Zamolodchikov and A.~B. Zamolodchikov, ``{Factorized s Matrices in
  Two-Dimensions as the Exact Solutions of Certain Relativistic Quantum Field
  Models},'' \href{http://dx.doi.org/10.1016/0003-4916(79)90391-9}{{\em Annals
  Phys.} {\bfseries 120} (1979) 253--291}.

\bibitem{He:2018uxa}
Y.~He, A.~Irrgang, and M.~Kruczenski, ``{A note on the S-matrix bootstrap for
  the 2d O(N) bosonic model},''
  \href{http://dx.doi.org/10.1007/JHEP11(2018)093}{{\em JHEP} {\bfseries 11}
  (2018) 093}, \href{http://arxiv.org/abs/1805.02812}{{\ttfamily
  arXiv:1805.02812 [hep-th]}}.

\bibitem{Cordova:2018uop}
L.~C\'ordova and P.~Vieira, ``{Adding flavour to the S-matrix bootstrap},''
  \href{http://dx.doi.org/10.1007/JHEP12(2018)063}{{\em JHEP} {\bfseries 12}
  (2018) 063}, \href{http://arxiv.org/abs/1805.11143}{{\ttfamily
  arXiv:1805.11143 [hep-th]}}.

\bibitem{Paulos:2018fym}
M.~F. Paulos and Z.~Zheng, ``{Bounding scattering of charged particles in $1+1$
  dimensions},'' \href{http://dx.doi.org/10.1007/JHEP05(2020)145}{{\em JHEP}
  {\bfseries 05} (2020) 145}, \href{http://arxiv.org/abs/1805.11429}{{\ttfamily
  arXiv:1805.11429 [hep-th]}}.

\bibitem{Cordova:2019lot}
L.~C\'ordova, Y.~He, M.~Kruczenski, and P.~Vieira, ``{The O(N) S-matrix
  Monolith},'' \href{http://dx.doi.org/10.1007/JHEP04(2020)142}{{\em JHEP}
  {\bfseries 04} (2020) 142}, \href{http://arxiv.org/abs/1909.06495}{{\ttfamily
  arXiv:1909.06495 [hep-th]}}.

\bibitem{Bercini:2019vme}
C.~Bercini, M.~Fabri, A.~Homrich, and P.~Vieira, ``{S-matrix bootstrap:
  Supersymmetry, $Z_2$, and $Z_4$ symmetry},''
  \href{http://dx.doi.org/10.1103/PhysRevD.101.045022}{{\em Phys. Rev. D}
  {\bfseries 101} no.~4, (2020) 045022},
  \href{http://arxiv.org/abs/1909.06453}{{\ttfamily arXiv:1909.06453
  [hep-th]}}.

\bibitem{Kruczenski:2020ujw}
M.~Kruczenski and H.~Murali, ``{The R-matrix bootstrap for the 2d O(N) bosonic
  model with a boundary},'' \href{http://arxiv.org/abs/2012.15576}{{\ttfamily
  arXiv:2012.15576 [hep-th]}}.

\bibitem{Guerrieri:2018uew}
A.~L. Guerrieri, J.~Penedones, and P.~Vieira, ``{Bootstrapping QCD Using Pion
  Scattering Amplitudes},''
  \href{http://dx.doi.org/10.1103/PhysRevLett.122.241604}{{\em Phys. Rev.
  Lett.} {\bfseries 122} no.~24, (2019) 241604},
  \href{http://arxiv.org/abs/1810.12849}{{\ttfamily arXiv:1810.12849
  [hep-th]}}.

\bibitem{Guerrieri:2020bto}
A.~Guerrieri, J.~Penedones, and P.~Vieira, ``{S-matrix Bootstrap for Effective
  Field Theories: Massless Pions},''
  \href{http://arxiv.org/abs/2011.02802}{{\ttfamily arXiv:2011.02802
  [hep-th]}}.

\bibitem{Doroud:2018szp}
N.~Doroud and J.~Elias~Mir\'o, ``{S-matrix bootstrap for resonances},''
  \href{http://dx.doi.org/10.1007/JHEP09(2018)052}{{\em JHEP} {\bfseries 09}
  (2018) 052}, \href{http://arxiv.org/abs/1804.04376}{{\ttfamily
  arXiv:1804.04376 [hep-th]}}.

\bibitem{EliasMiro:2019kyf}
J.~Elias~Mir\'o, A.~L. Guerrieri, A.~Hebbar, J.~a. Penedones, and P.~Vieira,
  ``{Flux Tube S-matrix Bootstrap},''
  \href{http://dx.doi.org/10.1103/PhysRevLett.123.221602}{{\em Phys. Rev.
  Lett.} {\bfseries 123} no.~22, (2019) 221602},
  \href{http://arxiv.org/abs/1906.08098}{{\ttfamily arXiv:1906.08098
  [hep-th]}}.

\bibitem{Karateev:2019ymz}
D.~Karateev, S.~Kuhn, and J.~a. Penedones, ``{Bootstrapping Massive Quantum
  Field Theories},'' \href{http://dx.doi.org/10.1007/JHEP07(2020)035}{{\em
  JHEP} {\bfseries 07} (2020) 035},
  \href{http://arxiv.org/abs/1912.08940}{{\ttfamily arXiv:1912.08940
  [hep-th]}}.

\bibitem{Correia:2020xtr}
M.~Correia, A.~Sever, and A.~Zhiboedov, ``{An Analytical Toolkit for the
  S-matrix Bootstrap},'' \href{http://arxiv.org/abs/2006.08221}{{\ttfamily
  arXiv:2006.08221 [hep-th]}}.

\bibitem{Bose:2020shm}
A.~Bose, P.~Haldar, A.~Sinha, P.~Sinha, and S.~S. Tiwari, ``{Relative entropy
  in scattering and the S-matrix bootstrap},''
  \href{http://dx.doi.org/10.21468/SciPostPhys.9.5.081}{{\em SciPost Phys.}
  {\bfseries 9} (2020) 081}, \href{http://arxiv.org/abs/2006.12213}{{\ttfamily
  arXiv:2006.12213 [hep-th]}}.

\bibitem{Komatsu:2020sag}
S.~Komatsu, M.~F. Paulos, B.~C. Van~Rees, and X.~Zhao, ``{Landau diagrams in
  AdS and S-matrices from conformal correlators},''
  \href{http://dx.doi.org/10.1007/JHEP11(2020)046}{{\em JHEP} {\bfseries 11}
  (2020) 046}, \href{http://arxiv.org/abs/2007.13745}{{\ttfamily
  arXiv:2007.13745 [hep-th]}}.

\bibitem{Bose:2020cod}
A.~Bose, A.~Sinha, and S.~S. Tiwari, ``{Selection rules for the S-Matrix
  bootstrap},'' \href{http://arxiv.org/abs/2011.07944}{{\ttfamily
  arXiv:2011.07944 [hep-th]}}.

\bibitem{Hebbar:2020ukp}
A.~Hebbar, D.~Karateev, and J.~Penedones, ``{Spinning S-matrix Bootstrap in
  4d},'' \href{http://arxiv.org/abs/2011.11708}{{\ttfamily arXiv:2011.11708
  [hep-th]}}.

\bibitem{Karateev:2020axc}
D.~Karateev, ``{Two-point Functions and Bootstrap Applications in Quantum Field
  Theories},'' \href{http://arxiv.org/abs/2012.08538}{{\ttfamily
  arXiv:2012.08538 [hep-th]}}.

\bibitem{Tourkine:2021fqh}
P.~Tourkine and A.~Zhiboedov, ``{Scattering from production in 2d},''
  \href{http://arxiv.org/abs/2101.05211}{{\ttfamily arXiv:2101.05211
  [hep-th]}}.

\bibitem{Guerrieri:2021ivu}
A.~Guerrieri, J.~Penedones, and P.~Vieira, ``{Where is String Theory?},''
  \href{http://arxiv.org/abs/2102.02847}{{\ttfamily arXiv:2102.02847
  [hep-th]}}.

\bibitem{Anderson:2016rcw}
P.~D. Anderson and M.~Kruczenski, ``{Loop Equations and bootstrap methods in
  the lattice},'' \href{http://dx.doi.org/10.1016/j.nuclphysb.2017.06.009}{{\em
  Nucl. Phys. B} {\bfseries 921} (2017) 702--726},
  \href{http://arxiv.org/abs/1612.08140}{{\ttfamily arXiv:1612.08140
  [hep-th]}}.

\bibitem{Elvang:2020lue}
H.~Elvang, ``{Bootstrap and Amplitudes: A Hike in the Landscape of Quantum
  Field Theory},'' \href{http://arxiv.org/abs/2007.08436}{{\ttfamily
  arXiv:2007.08436 [hep-th]}}.

\bibitem{Huang:2020nqy}
Y.-t. Huang, J.-Y. Liu, L.~Rodina, and Y.~Wang, ``{Carving out the Space of
  Open-String S-matrix},'' \href{http://arxiv.org/abs/2008.02293}{{\ttfamily
  arXiv:2008.02293 [hep-th]}}.

\bibitem{Guerrieri:2020kcs}
A.~L. Guerrieri, A.~Homrich, and P.~Vieira, ``{Dual S-matrix bootstrap. Part I.
  2D theory},'' \href{http://dx.doi.org/10.1007/JHEP11(2020)084}{{\em JHEP}
  {\bfseries 11} (2020) 084}, \href{http://arxiv.org/abs/2008.02770}{{\ttfamily
  arXiv:2008.02770 [hep-th]}}.

\bibitem{pollica}
L.~F. Alday {\em et~al.}, ``The pollica perspective on the (super)-conformal
  world,'' {\em Journal of Physics A: Mathematical and Theoretical} (2021) .
  \url{http://iopscience.iop.org/article/10.1088/1751-8121/abf38e}.

\bibitem{LM}
C.~Lopez and G.~Mennessier, ``{Bounds on the pi0 pi0 Amplitude},''
  \href{http://dx.doi.org/10.1016/0550-3213(77)90237-1}{{\em Nucl. Phys. B}
  {\bfseries 118} (1977) 426--444}.

\bibitem{conic1}
Z.-Q. Luo, J.~Sturm, and S.~Zhang, ``{Duality and Self-Duality for Conic Convex
  Programming},'' Econometric Institute Research Papers EI 9620-/A, Erasmus
  University Rotterdam, Erasmus School of Economics (ESE), Econometric
  Institute, Jan., 1996.
\newblock \url{https://ideas.repec.org/p/ems/eureir/1381.html}.

\bibitem{cvx}
M.~Grant and S.~Boyd, ``{CVX}: Matlab software for disciplined convex
  programming, version 2.1.'' \url{http://cvxr.com/cvx}, Mar., 2014.

\bibitem{gb08}
M.~Grant and S.~Boyd, ``Graph implementations for nonsmooth convex programs,''
  in {\em Recent Advances in Learning and Control}, V.~Blondel, S.~Boyd, and
  H.~Kimura, eds., Lecture Notes in Control and Information Sciences,
  pp.~95--110.
\newblock Springer-Verlag Limited, 2008.

\bibitem{boyd2004convex}
S.~Boyd, L.~Vandenberghe, and C.~U. Press, {\em Convex Optimization}.
\newblock No.~pt. 1 in Berichte {\"u}ber verteilte messysteme. Cambridge
  University Press, 2004.

\bibitem{groetsch1984theory}
C.~Groetsch, {\em The Theory of Tikhonov Regularization for Fredholm Equations
  of the First Kind}.
\newblock Chapman \& Hall/CRC research notes in mathematics series.

\bibitem{duren1970}
\href{http://dx.doi.org/https://doi.org/10.1016/S0079-8169(08)62672-0}{``Chapter
  8 - extremal problems,''} in {\em Theory of Hp Spaces}, P.~L. Duren, ed.,
  vol.~38 of {\em Pure and Applied Mathematics}, pp.~129--146.
\newblock Elsevier, 1970.

\bibitem{gradshteyn2007}
I.~S. Gradshteyn and I.~M. Ryzhik, {\em Table of integrals, series, and
  products}.
\newblock Elsevier/Academic Press, Amsterdam, seventh~ed., 2007.

\end{thebibliography}\endgroup

\end{document}